\documentclass[fleqn,usenatbib]{mnras}

\usepackage[T1]{fontenc}
\usepackage{ae,aecompl}
\usepackage{geometry}
\usepackage{amsmath}
\usepackage{multicol}
\usepackage{graphicx}
\usepackage{siunitx}
\usepackage{subfig}
\usepackage{amssymb}	
\usepackage{cleveref}
\usepackage[normalem]{ulem}

\graphicspath{{./}{figures/}}

\newcommand{\avg}[1]{\left\langle #1 \right\rangle}
\newcommand\numberthis{\addtocounter{equation}{1}\tag{\theequation}}
\DeclareSIUnit \parsec {pc}
\DeclareSIUnit \jansky {Jy}

\newcommand{\tr}{\text{Tr}}

\newcommand{\mat}[1]{\mathrm{#1}}
\newcommand{\vecb}[1]{\boldsymbol{#1}}

\crefname{figure}{Fig.}{Figs.}
\Crefname{figure}{Fig.}{Figs.}

\title[Fundamental uncertainty levels of 21cm power spectra from a delay analysis]{Fundamental uncertainty levels of 21cm power spectra from a delay analysis}

\author[A. E. Lanman et al.]{
Adam E. Lanman,$^{1}$\thanks{E-mail: adam\_lanman@brown.edu}
Jonathan C. Pober$^{1}$
\\
$^{1}$Brown University, Department of Physics
}

\pubyear{2019}

\begin{document}
\label{firstpage}
\pagerange{\pageref{firstpage}--\pageref{lastpage}}
\maketitle

\begin{abstract}
Several experimental efforts are underway to measure the power spectrum of 21cm fluctuations from the Epoch of Reionization (EoR) using low-frequency radio interferometers. Experiments like the Hydrogen Epoch of Reionization Array (HERA) and Murchison Widefield Array Phase II (MWA) feature highly-redundant antenna layouts, building sensitivity through redundant measurements of the same angular Fourier modes, at the expense of diminished UV coverage. This strategy limits the numbers of independent samples of each power spectrum mode, thereby increasing the effect of sample variance on the final power spectrum uncertainty. To better quantify this effect, we measure the sample variance of a delay-transform based power spectrum estimator, using both analytic calculations and simulations of flat-spectrum EoR-like signals. We find that for the shortest baselines in HERA, the sample variance can reach as high as 20\%, and up to 30\% for the wider fields-of-view of the MWA. Combining estimates from all the baselines in a HERA- or MWA-like 37 element redundant hexagonal array can lower the variance to $1-3$\% for some Fourier modes. These results have important implications for observing and analysis strategies, and suggest that sample variance can be non-negligible when constraining EoR model parameters from upcoming 21cm data.
\end{abstract}


\section{Introduction}
\label{sec:intro}

The Epoch of Reionization (EoR) was the last global phase transition of the universe, when the first UV emitting sources formed and carved out bubbles of ionization in the previously-neutral intergalactic medium. The lack of bright sources and abundance of absorbing gas makes the EoR, and the dark ages that preceded it, inaccessible to traditional high redshift survey techniques. The hyperfine ``spin-flip'' transition of neutral hydrogen, which emits a photon of rest-wavelength 21cm, offers one of the most promising observational probes of the EoR \citep{furlanetto_cosmology_2006, morales_reionization_2010}. The brightness of 21cm emission/absorption is tightly bound to the ionization state, density, and temperature of the gas, and as a line transition at cosmological distances, the redshift directly maps to distance. This technique therefore has the potential to directly map the three-dimensional structure of hydrogen during the EoR. The low opacity of hydrogen to 21cm means that this signal is both extremely weak and unobstructed by intervening gas, with brightness contrasts expected to be of order \si{\milli\kelvin} \citep{pritchard_evolution_2008}

The wealth of information promised by 21cm observations has spurred a renewed interest in low-frequency radio astronomy over the last decade. However, the expected weakness of the signal presents a considerable challenge. Directly imaging structures of the EoR will likely require an instrument of the scale of the upcoming Square Kilometre Array (SKA; \citealt{mellema_reionization_2013, furlanetto_cosmology_2006}), so the current generation instruments are seeking statistical measures of the EoR signal, especially the power spectrum. Experiments like the GMRT \citep{paciga_simulation-calibrated_2013, paciga_gmrt_2011}, MWA \citep{beardsley_first_2016, bowman_science_2013}, PAPER \citep{parsons_precision_2010, ali_paper-64_2015, cheng_characterizing_2018}, and LOFAR \citep{patil_upper_2017}, have placed upper limits on the power spectrum amplitude at several redshifts.


Choosing the optimal interferometer design to maximize power spectrum sensitivity requires balancing the effects of antenna positions, receiver element, and computational requirements of the design. The most significant factors that need to be reduced are thermal noise and foreground power \citep{franzen_154_2016, yatawatta_initial_2013, bernardi_foregrounds_2010, jelic_foreground_2008}, so much work has focused on these effects. Short baselines are more sensitive to diffuse structures like the EoR signal, and a random antenna distribution typically improves imaging capability, which helps foreground modeling and subtraction \citep{lidz_detecting_2008}. This motivated the Phase I layout of the Murchison Widefield Array (MWA), in Murchison Radio Observatory in Western Australia, which consisted of a random distribution of 128 elements, featuring a dense core but with several baselines reaching as far as a few kilometers. On the other hand, a highly-redundant array can be made more sensitive to line of sight Fourier modes with relatively few antennas, because redundant baselines directly sample the same angular Fourier modes ($k_\perp$) \citep{dillon_redundant_2016, parsons_sensitivity_2012}. This has the additional benefit of improving calibration through redundant techniques \citep{liu_precision_2010}.

With these considerations, many current EoR experiments feature highly redundant layouts, fine spectral resolution, and drift-scanning, zenith-pointing antennas. The Hydrogen Epoch of Reionization Array (HERA), under construction in the Karoo desert of South Africa, comprises a densely packed hexagonal grid of 14.6\si{\meter} parabolic dishes with dual-polarization feeds at the prime focus \citep{deboer_hydrogen_2017}. It is sensitive to a bandpass of 100 -- 200 \si{\mega\hertz} with 1024 frequency channels (97 \si{\kilo\hertz} channel width). When finished, it will have 350 dishes, 30 of which are outriggers placed farther away from the compact core to improve imaging capabilities.

The Phase II deployment of the MWA is a hybrid design \citep{wayth_phase_2018} with 54 of the original core tiles of Phase I left in place, and the remaining 74 tiles re-purposed into two 37 element hexagonal grids with a 14 \si{\meter} spacing. Each ``tile'' is a phased array of 16 dual-polarization dipole antennas acting as a single element in the correlator. This hybrid design allows for a combination of calibration and foreground removal techniques to be applied, and provides data with well-tested hardware for comparison with HERA.

The main trade-off to these redundant designs is that for more antennas there are fewer independent samples of each Fourier mode, due to the redundancy of many measurements. Statistical quantities such as the power spectrum have an intrinsic uncertainty, called \emph{sample or cosmic variance}, due to the limited size of the volume surveyed by an instrument. For a Gaussian signal, this is proportional to the power spectrum amplitude over the number of independent samples $N$ (see \citealt{mcquinn_cosmological_2006}). This holds true for non-Gaussian signals when $N$ is sufficiently large, due to the central limit theorem. The root-mean-square (RMS) error is the square root of the sample variance, and drops off as $1/\sqrt{N}$. Assuming no correlation between neighboring times, the number of independent samples of each $k_\parallel$ mode is just the number of integration times, but real observations do have correlations in time. Each radio interferometric measurement (visibility) is an integral over the primary beam across the sky, and visibilities at neighboring times will integrate over overlapping regions. The effective number of independent samples in a given survey volume is decreased by this correlation, so the sample variance decreases more slowly.

Previous sensitivity estimates have argued that sample variance is negligible except in a handful of k-modes, and so subsequent analysis has mostly ignored the issue in light of the more significant issues of thermal noise and foreground contamination. Since foreground sources tend to be smooth spectrum, it is possible to isolate foreground power to low-$k_\parallel$ modes, since $k_\parallel$ is proportional to the Fourier dual of frequency. This power tends to spread to higher $k_\parallel$ modes for longer baselines, forming the so-called \emph{foreground wedge} \citep{datta_bright_2010, morales_four_2012, hazelton_fundamental_2013, thyagarajan_foregrounds_2015}. For this reason, many published power spectrum limits use data from short baselines only. For example, the power spectrum limits from PAPER \citep{ali_paper-64_2015, cheng_characterizing_2018} were made using only the three shortest baselines in the array.

Ultimately, measurements of 21cm power spectra from the cosmic dawn will be used to constrain cosmological and astrophysical model parameters using such methods as Markov Chain Monte Carlo (MCMC) \citep{greig_21cmmc:_2015} and machine learning \citep{schmit_emulation_2017, gillet_deep_2019}. \cite{greig_21cmmc:_2015} found that a 25\% uncertainty in the measured power spectrum can increase uncertainty in model parameter estimates by a factor of a few, significantly degrading the constraints on the properties of galaxies during the EoR (although sampling the 21cm power spectrum over a wide range of redshifts can reduce the effect). Such uncertainty must be understood and accounted for in the era of precision cosmology.

In this paper, we argue that the sample variance of single-baseline power spectrum estimation does not fall below this 25\% threshold for a typical data volume. In other words: even if all foreground and noise power can be properly mitigated, the fundamental uncertainty of the estimated 21cm power spectrum will be too high for it to be used for precision cosmology. Through instrument simulations of a mock EoR signal, we explore how this sample variance is affected by baseline length and beam width, and how it can be mitigated by combining data from multiple non-redundant baselines.

\section{Power Spectrum Sensitivity}

Many previous measures of instrument sensitivity estimated the power in Fourier space of thermal noise and residual foreground contamination, and made simplified approximations of sample independence to calculate the variance in each Fourier mode \citep[e.g.][]{beardsley_eor_2013,pober_what_2014}. The approximate integration time per Fourier mode is done by dividing the UV plane into bins of size given by the effective antenna aperture in wavelengths and calculating the motion of baselines with time and frequency. The variance is estimated by modeling thermal noise and residual foreground power and dividing through by this integration time. Sample variance is estimated as a component proportional to the expected power spectrum amplitude.

\cite{beardsley_eor_2013} used a method like this to predict that the MWA Phase I would be capable of detecting the EoR power spectrum, but relied on optimistic assumptions as to the degree which residual foreground power contaminated the signal \citep{dillon_empirical_2015}. \cite{pober_what_2014} estimated the sensitivity of HERA in this way by combining the thermal noise uncertainty, from a formula presented in \cite{pober_baryon_2013}, and residual foreground power from three foreground models.

\cite{thyagarajan_study_2013} considered the contributions of thermal noise, residual foreground power, and sample variance to the uncertainty of MWA Phase I observations in comparing two observing modes: a long integration on a single field, or (for an equal total observing time) fewer observations of multiple fields.  As a phased array, each MWA tile is capable of pointing in discrete directions. This allows for a ``drift and shift'' observing strategy, where the array can approximately track a field by phasing to a position just west of it, letting the field drift through, then ``shifting'' the pointing west again. \cite{thyagarajan_study_2013} also took into account a more realistic model of the antenna power pattern (beam) in UV space, providing a more realistic look at how primary beam effects (especially sidelobe sources) contribute to foreground power leakage. Sample variance was assumed to drop off as one over the number of fields, i.e., for perfectly independent fields.

\cite{trott_comparison_2014} performed a similar analysis for MWA Phase I, but instead of looking at discrete fields also directly considered the effects of covariance between visibilities in a constantly-evolving \emph{drift scan}. In this case, the independence of observing fields is harder to determine. \cite{trott_comparison_2014} found that the choice of ideal observing strategy depended on the available observing time and survey volume. The drift-scanning mode reduced sample variance the most for a given observing time, as one might expect since drift-scanning covers a larger area, but did not reduce thermal noise as significantly. For a drift scanning approach, they estimate a best signal to noise ratio of 9.6 (or an RMS error of about 10\%) for the power spectrum amplitude.

This paper takes an exact approach to study the effects of sample variance on realistic 21cm observations for redundant arrays. By simulating the observations of an idealized interferometer over 24 hours of observing time, we can directly measure the variance of the measured power spectra and the covariance of delay-transformed visibilities. We simulate observations of a full sky model with a known power spectrum, and measure the sample variance of power spectra estimated from the simulated datasets.

We ignore the effects of foregrounds and thermal noise in this work, assuming that foreground power has been sufficiently removed through sky modeling techniques. Thermal noise can in principle be reduced as much as desired by including data from multiple days. The sample variance ultimately is the most fundamental uncertainty, because we are limited to one sky.

We introduce our simplified delay spectrum estimator, similar to that used by PAPER and will be used by HERA, and discuss its expected statistics in \cref{sec:delay_spec}. \Cref{sec:toy_model} further explores the statistics of the estimator through controlled numerical tests, examining how visibility correlation relates to the distribution of estimator values. We describe our full simulation code and sky model in \cref{sec:simulation}. \Cref{sec:settings} describes our main questions and the simulations used to address them. In \cref{sec:results} we present the results of several sets of simulations, exploring the correlations of visibilities with time and the sample variance of power spectrum estimates.

\section{The Delay Spectrum}
\label{sec:delay_spec}


The delay transform was introduced in \citep{parsons_calibration_2009, parsons_per-baseline_2012, parsons_new_2014} as an inverse Fourier transform along the frequency axis into ``delay'' space.
\begin{align}
\tilde{V} (\vecb{b}, \tau, t) &= \int V(\vecb{b}, \nu, t) e^{2\pi i \nu \tau } d\nu\\
&= \int \int A(\hat{s}) \: T(\hat{s}, \nu, t) e^{-2 \pi i (\vecb{b} \cdot \hat{s} \nu /c - \nu \tau) } d\Omega d\nu
\label{eqn:delay_rime}
\end{align}
\Cref{eqn:delay_rime} gives the delay-transform of the visibility measured by baseline $\vecb{b}$, with primary antenna beam $A$, at time $t$. $T(\hat{s}, \nu, t)$ is the fluctuating component of the sky brightness temperature at frequency $\nu$ which has mean 0. \Cref{eqn:delay_rime} is explicitly written here in instrument (topocentric) coordinates, such that the sky moves while the primary beam and fringe terms are fixed in time. The unit vector $\hat{s}$ represents a position on the sky in this topocentric frame.

Delay modes $\tau$ approximately correspond with Fourier modes parallel to the line of sight $k_\parallel$, while the baseline length $|\vecb{u}|$ corresponds to perpendicular modes $k_\perp$. We can therefore estimate the power spectrum by cross-multiplying delay-transformed visibilities for redundant baselines \citep{parsons_new_2014}.
\begin{align*}
\widehat{P}(k_{b,\tau},t) &\equiv \left(\frac{\lambda^2}{2 k_B} \right)^2 \frac{X^2 Y}{B \Omega_\text{pp}} \avg{ \tilde{V}_\alpha (\tau, t) \tilde{V}^*_\beta (\tau, t)}_{\alpha,\beta \in G_{b}} \numberthis \label{eqn:pk_full} \\
k_{b,\tau} &= 2 \pi \sqrt{\left(\frac{\tau}{Y}\right)^2 + \left(\frac{b}{\lambda X}\right)^2 }
\end{align*}
The Fourier mode $k_{b,\tau}$ probed by this estimator depends on the baseline length $u$ (in wavelengths) as well as the delay mode $\tau$. An average is taken over baselines $\alpha,\beta$ that are in the same redundant group $G$. $X$ and $Y$ are a cosmological scaling factors with units of length per angle and length per frequency, respectively. $\lambda$ is the observed wavelength and $k_B$ is Boltzmann's constant. $\Omega_\text{pp}$ is the integral of the primary beam squared, and $B$ is the bandwidth of the observation (see Appendix B of \citealt{parsons_new_2014}).

Cosmological isotropy requires that the power spectrum be independent of direction, allowing us to get a better estimate by averaging samples in time:
\begin{equation}
\widehat{P}_\text{avg}(k_{b,\tau}; N_t) = \frac{1}{N_t} \sum\limits_{i=0}^{N_t} \widehat{P}(\vecb{k}_\tau,t_i)
\label{eqn:pk_avg}
\end{equation}
where $t_i$ are the separate integration times of the instrument, of which we're averaging together delay spectra from $N_t$ time samples.

For a single baseline ($\alpha = \beta$), we can combine \cref{eqn:pk_full,eqn:pk_avg} into a single expression.
\begin{equation}
	\widehat{P}(k_{b,\tau}, N_t; b) = \frac{\Phi}{N_t} \sum\limits_{n = 1}^{N_t} |\widetilde{V}(b, \tau,t_n)|^2
	\label{eqn:pspec_est}
\end{equation}
where $\Phi = (\lambda^2/2k_B)^2 X^2Y/B \Omega_{pp}$ is a scalar combining the cosmological and volumetric scaling factors of \cref{eqn:pk_full}. Note that cross-multiplying data for a single baseline will lead to a bias in the presence of thermal noise. Cross-multiplying redundant baselines, which measure the same signal but have independent noise realizations, avoids this bias. We can also avoid this bias by cross-multiplying measurements of the same LSTs across multiple nights. The analysis presesnted in this work doesn't include thermal noise, so we can use \cref{eqn:pspec_est} as an unbiased estimator.

\subsection{Statistics of $\widehat{P}(k)$}

The delay-transformed visibilities may be written as a measurement vector $\vecb{v}_\tau = (v_0, \ldots, v_N)$, with $v_n = \widetilde{V}(\tau, t_n)$. For a Gaussian sky signal, $\vecb{v}_\tau$ follows a complex multivariate gaussian distribution, with mean vector $\vecb{\mu}=0$ and covariance matrix $\mat{C} = \avg{\vecb{v}_\tau \vecb{v}_\tau^\dagger}$. $\mat{C}$ is an $N_t\times N_t$ symmetric, real, positive semi-definite matrix.

We may write our estimator \eqref{eqn:pspec_est} as
\begin{equation}
    \widehat{P}_N \equiv \widehat{P}(k_{b,\tau}, N_t; b) = \vecb{v}_\tau^\dagger \mat{A} \vecb{v}_\tau
\end{equation}
where $\mat{A} = (\Phi/N_t) \mathcal{I}_N$ for $(N_t\times N_t)$ identity matrix $\mathcal{I}_N$. We let $\widehat{P}_N$ stand for \cref{eqn:pspec_est} to simplify notation, with the understanding that it refers to a single baseline and delay mode, likewise dropping the $\tau$ subscript on $\vecb{v}$. In this notation, $\widehat{P}_N$ is recognizable as a real quadratic form. In the case that visibilities are uncorrelated and have similar noise, eq. \eqref{eqn:pspec_est} follows a $\chi^2$ distribution with $N_t$ degrees of freedom. More generally, this quadratic form may be shown to be $\chi^2$ distributed with degrees of freedom $d = \text{rank}(\mat{AC})$ under the condition that the matrix $\mat{A} \mat{C}$ is idempotent \citep{mathai1992quadratic, searle_linear_models_1971}. We find experimentally that \cref{eqn:pspec_est} is well-described by a $\chi^2$ distribution (see \cref{sec:toy_model}), despite a lack of idempotency of $\mat{AC}$. The underlying distribution of \cref{eqn:pspec_est} is not important to our results, but is of interest for interpreting the measured uncertainty. We will fit $\chi^2$ distributions to the measured power spectra.

The variance of the estimator may be found directly from the covariance matrix using Theorem 1 of Chapter 5 of \cite{searle_linear_models_1971}. The proof presented there readily extends to the complex case with the result:
\begin{equation}
    \text{Var}[\vecb{v}^\dagger \mat{A} \vecb{v}]  = \tr[(\mat{A C})^\dagger (\mat{A C})] + 2 \vecb{\mu}^\dagger \mat{A C A} \vecb{\mu}
\end{equation}

For the delay-transformed visibilities, $\vecb{\mu} = 0$, so this expression reduces to the trace of the squared covariance matrix with the scalar factors. Recalling that $C_{ij}$ is Hermitian, it follows that the variance becomes the sum of the squares of all entries:
\begin{equation}
\text{Var}[\widehat{P}_{N_t}] = \left(\frac{\Phi}{N}\right)^2 \sum\limits_{i,j}^{N_t} |\mat{C}_{ij}|^2
\label{eqn:varpk_trace}
\end{equation}

For the case of uncorrelated visibilities with variances $\sigma^2$, this reduces to $\text{Var}[\widehat{P}_N] = \Phi^2 \sigma^2 /N$, as expected. To understand the variance of our estimator, therefore, we need to look at the covariance between visibility measurements at different times.

\subsection{Visibility covariance}
\label{sec:vis_covar}

\begin{figure*}
    \centering
    \subfloat[\label{fig:covmat}]{\includegraphics[width=0.5\textwidth]{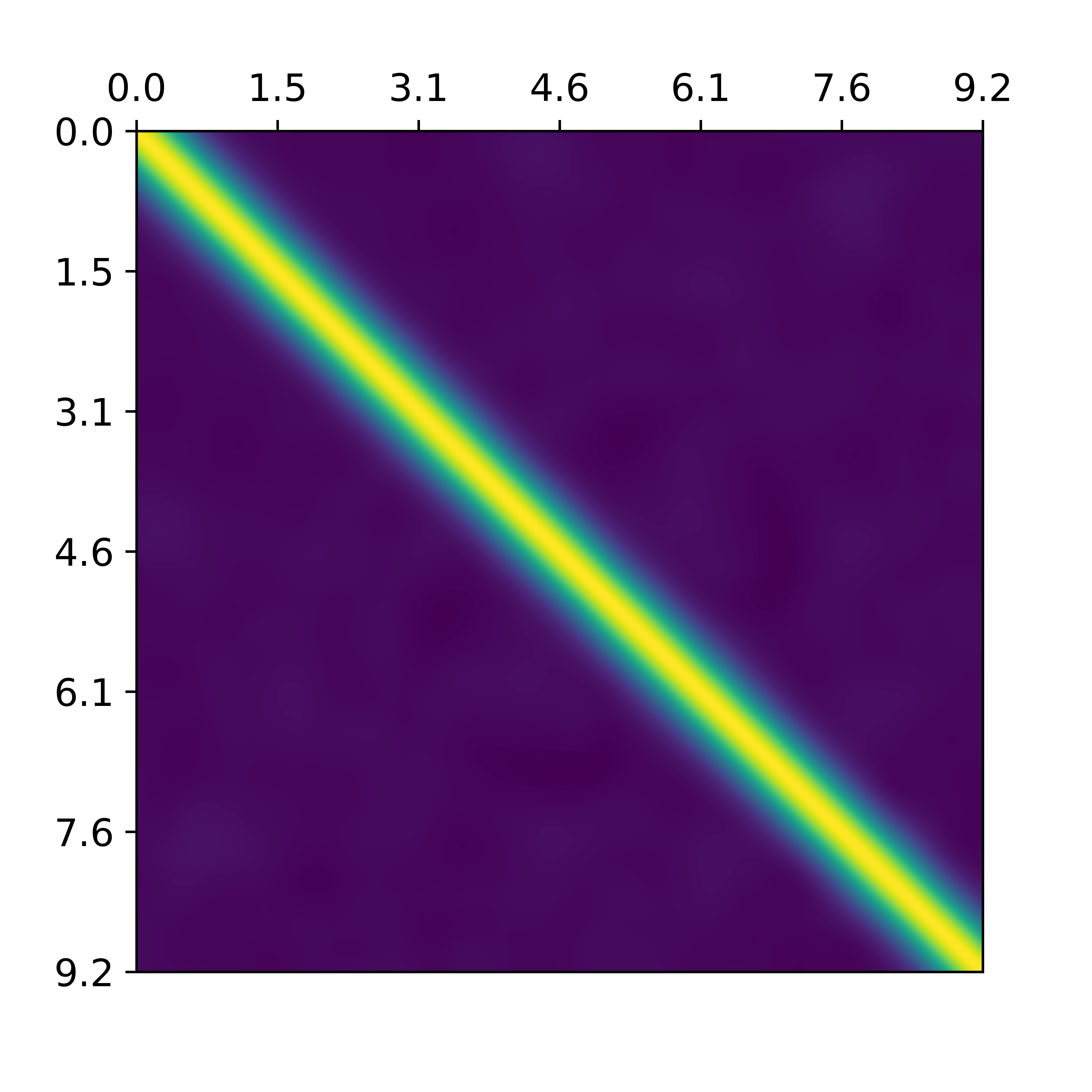}}
    \subfloat[\label{fig:corfunc}]{\includegraphics[width=0.5\textwidth]{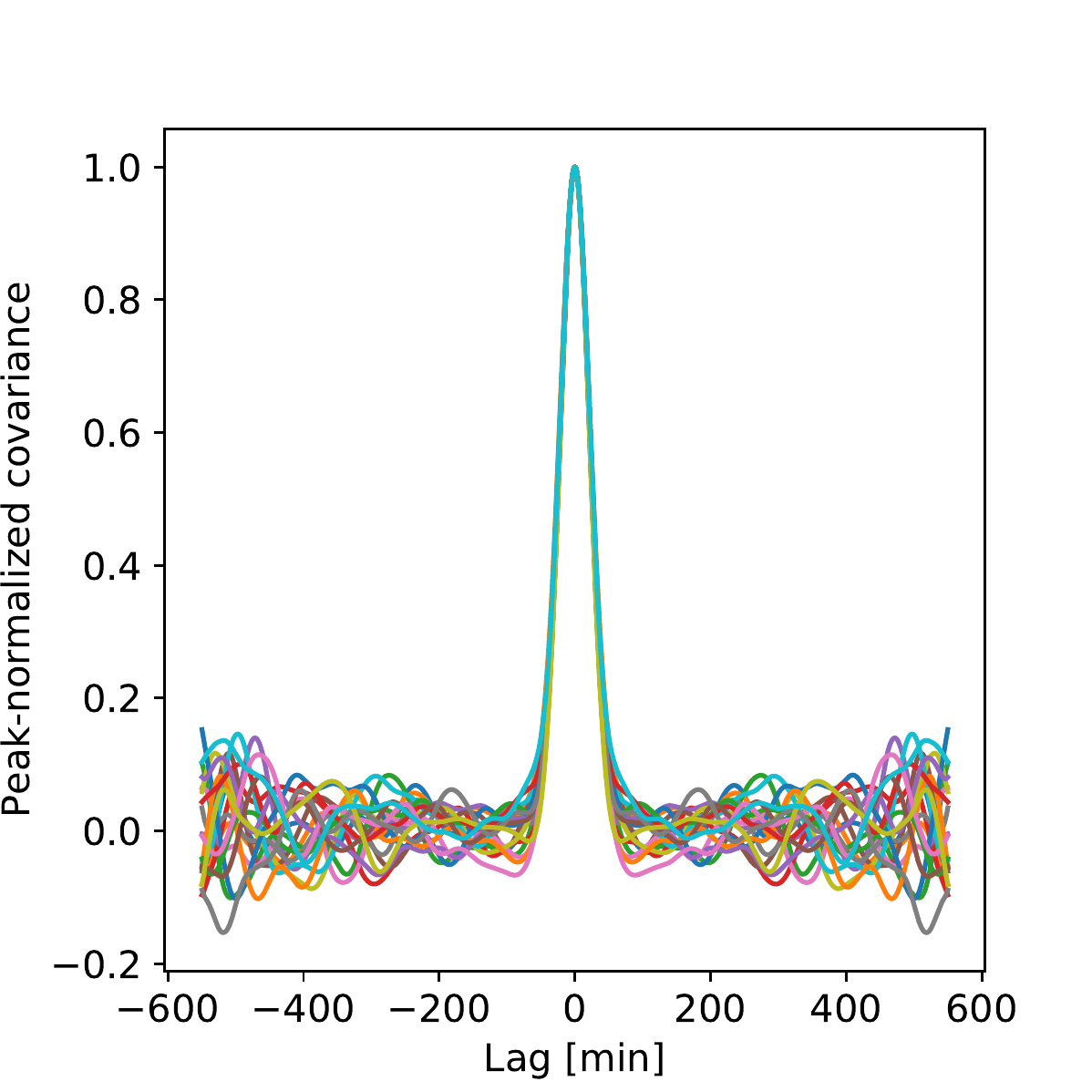}}
    \caption{\protect\subref{fig:covmat} Covariance function averaged over delay, for a single baseline, from simulated data. Both axes are time in hours. \protect\subref{fig:corfunc}  The measured correlation function, made from binning the covariance matrix in lag for each delay mode, scaled to 1 at a lag of 0. There are fewer entries in the covariance matrix that correspond with the large lags, which leads to the increased scatter away from the central peak. The central peak is well-described by the gaussian function.}
    \label{fig:example_covariance}
\end{figure*}

A full expression for the covariance between two delay-transformed visibilities has been worked out in \cite{zhang_unlocking_2018} in detail. For a single baseline and drift-scanning visibilities, their expression reduces to the form
\begin{alignat*}{1}
\mat{C}_{ij} = \frac{P(k_{b,\tau})}{X^2 Y} \left(\frac{2 k_B}{\lambda^2}\right)^2& \int \int A^*(\hat{s}, \nu) A(\mat{R}_{ij} \hat{s}, \nu) \\
&\times \exp \left[ 2 \pi i \frac{\nu}{c} (\hat{s} - \mat{R}_{ij}\hat{s}) \cdot \vecb{b} \right]  d\Omega\: d\nu  \numberthis
\label{eqn:delay_vis_covar} \\
\end{alignat*}
\Cref{eqn:delay_vis_covar} assumes the power spectrum is approximately constant across the bandpass (i.e. ignoring light-cone effects). $\mat{R}_{ij}$ is a three-dimensional rotation matrix describing the motion of source from time $t_i$ to $t_j$ in the observer's topocentric frame (e.g., altitude/azimuth).\footnote{A notational note -- The subscripts on $\mat{R}_{ij}$ are not indexes of the matrix itself.} Note that in the case of $t_i=t_j$, $\mat{R}_{ij} = \mathcal{I}$ and we recover the estimator \eqref{eqn:pk_full} from \cref{eqn:delay_vis_covar}.

We can manipulate \cref{eqn:delay_vis_covar} further to gain some better intuition. Let $\Theta_\nu$ represent the sky integral, such that,
$$
C_{ij} = \int d\nu \Theta_\nu
$$

Then
\begin{alignat}{2}
\Theta_\nu &= \int d\Omega A^*(\hat{s}) A(R_{ij}\hat{s}) \exp \left[
-2 \pi i \frac{\nu}{c} ( \mathcal{I} - \mat{R}_{ij}) \hat{s} \cdot \vecb{b}
 \right] \nonumber \\
&= \int d\Omega A_\nu^*(\hat{s}) A(\mat{R}_{ij}\hat{s}) \exp \left[ -2 \pi i \frac{\nu}{c}\hat{s} \cdot \vecb{b}'\right] \label{eqn:delay_corr_sky}\\
&\qquad \text{where } \vecb{b}' = (\mathcal{I} - \mat{R}_{ij}^T) \vecb{b}\nonumber
\end{alignat}

In this form, we can see that $\Theta_\nu$ is given by the overlap of the beam between the pointings, weighted by a fringe term. The fringe width is given by length of the baseline projected onto the displacement vector of the sky, $|\vecb{b}'|$.

If we choose a fixed phase center $\hat{s}_0$, we can rewrite \cref{eqn:delay_corr_sky} under a flat-sky approximation. We define $\vec{b}'$ as the projection of $\vecb{b}'$ onto the plane orthogonal to $\hat{s}_0$, and $\vec{l} = \hat{s} - (\hat{s} \cdot \hat{s}_0) \hat{s}_0$ is the projection of the sky position vector into the sky plane.\footnote{We denote 2D vectors with overhead arrows.} Finally, we denote by $\mat{\Gamma}_{ij}$ the transformation matrix of $\vec{l}$ corresponding with the 3D transformation $R_{ij}$.

In these terms, \cref{eqn:delay_corr_sky} may be written
\begin{equation}
    \Theta_\nu = \int A_\nu^*(\vec{l}) A_\nu(\mat{\Gamma}_{ij} \vec{l})
    \exp\left[ -2\pi i \frac{\nu}{c} \vec{l} \cdot \vec{b}' \right]
\end{equation}

In this form, the covariance is a 2D Fourier transform of a product of the primary beams, which may be expressed as the convolution of the Fourier-transformed primary beams, which we call \emph{beam kernels}:

\begin{equation}
\Theta_\nu \simeq \int\widetilde{A}^*\left(\frac{\nu}{c} \vec{b}' - \vec{u} \right) \widetilde{A} (\mat{\Gamma}_{ij}^T \vec{u}) \: d^2 u
\label{eqn:fourier_corr}
\end{equation}

This quantity depends only on the baseline, beam kernel width, and time between visibilities (\emph{lag}). Since beam kernels are compact in Fourier space, the correlation is likewise limited in range, and can be characterized by a \emph{correlation time} $t_c$. For our purposes, we define $t_c$ to be the full-width at half maximum of the correlation vs. lag. Note that in the case of nearly-Gaussian primary beam functions, the convolution in \cref{eqn:fourier_corr} will give a Gaussian function, which can be characterized entirely by its FWHM.

The width of beam kernel, typically equal to the antenna size in wavelengths, scales inversely with the primary beam width on the sky. From \cref{eqn:fourier_corr} we therefore expect that measurements from shorter baselines and narrower beams should stay correlated longer than those from long baselines and wide beams. This may seem counter-intuitive, so it's worth some further discussion. One may think of the primary (antenna) beam as selecting a ``patch'' of sky that is integrated into a single visibility. In this picture, visibilities which sample overlapping parts of the sky should be correlated, and so the correlation time should be the time it takes the sky to pass through the primary beam. However, the exponential term in \cref{eqn:delay_corr_sky} complicates this interpretation, and makes it more sensible to think about the Fourier-transformed formula \cref{eqn:fourier_corr}, which is more in line with what previous sensitivity analyses have done. We will show in \cref{sec:cov_meas}, however, that the Fourier-space interpretation of the correlation is not always valid.

\Cref{fig:covmat} shows an example covariance matrix obtained from simulated data. The simulations used to make this data are discussed in \cref{sec:simulation}. This is the correlation in time for a single baseline, averaged over delay, and shows the expected symmetry due to the isotropy of the sky signal. If we bin the covariance matrices for each delay mode in time separation, or \emph{lag}, we get \cref{fig:corfunc}, which we call the \emph{correlation function}. This shows that the central stripe in the covariance matrix has a nearly Gaussian shape.

\subsection{Theoretical sample variance}

We can construct an analytic function for the variance, \cref{eqn:varpk_trace}, under some assumptions for the visibility correlation. We first assume that the time covariance matrix is symmetric and constant parallel to the diagonal (also known as a symmetric Toeplitz matrix). This matrix symmetry is equivalent to the assumption that the EoR signal is isotropic since the correlation matrix depends only on the separation in time index,
$$
\gamma_n = \Phi^2 |C_{ij}|^2 \qquad \text{ such that  } n = |i - j|
$$
The factor of $\Phi$, from \cref{eqn:pspec_est}, is included to absorb the cosmological and unit scalars into the definition of $\gamma_n$. We can express the sum in \cref{eqn:varpk_trace} by counting the number of matrix entries at each $n$.
\begin{equation}
\text{Var}[P_k] = \frac{1}{N}\left( 2 \sum\limits_{n=0}^{N - 1} \gamma_n   -\gamma_0 \right) - \frac{2}{N^2} \sum\limits_{n=0}^{N-1} n \gamma_n
\label{eqn:var_symm}
\end{equation}

To continue, we will treat $\gamma_n$ as a discretization of a continuous correlation function $\gamma(t)$. Assuming the time is discretized into time steps of length $\delta t$, such that $t_n = n \Delta t$, with a total time $T = N \delta t$, the sums in \cref{eqn:var_symm} become:
\begin{align}
    \frac{1}{N}\sum\limits_{n=0}^{N-1} \gamma_n & 
    \approx \frac{1}{N\Delta t} \int\limits_0 \gamma(t) dt = \frac{1}{T}\int\limits_0^T \gamma(t) dt\\
    \frac{1}{N^2}\sum\limits_{n=0}^{N-1} n\gamma_n &
    \approx \frac{1}{N^2 \Delta t^2} \int\limits_0 t  \gamma(t) dt = \frac{1}{T^2}\int\limits_0^T t \gamma(t) dt
\end{align}

Approximating the correlation function as a Gaussian with FWHM $t_c$, we can evaluate these integrals. Letting $w_c = t_c/2.355$, the correlation function is 
\begin{equation}
    \gamma(t) = \gamma_0 \exp\left(\frac{-t^2}{2 w_c^2}\right)
\end{equation}

Carrying out the integrals in \cref{eqn:var_symm} with this Gaussian, we're left with the following for the variance of the power spectrum estimator at averaging time $T$:
\begin{align}
    \text{Var}[\widehat{P}_k(T)] = \gamma_0^2 &\left( \frac{w_c}{T} \sqrt{2 \pi} \: \text{Erf}\left[\frac{T}{\sqrt{2}w_c}\right] \right.\nonumber\\
     &- \left. 2 \left(\frac{w_c}{T}\right)^2 \left[ 1 - e^{-T^2 / 2 w_c^2}\right]
    \right)
    \label{eqn:variance_function}
\end{align}
Erf$[\cdot]$ is the error function. This expression holds only under the assumption that $\Delta t \ll T$. In \cref{sec:discussion} we show that this approximate expression fits with the sample variances measured from simulated data.

\section{Toy Model}
\label{sec:toy_model}

To better understand the relationship between the distribution of $\widehat{P}_N$ and the correlation of visibility data, we ran a series of simple tests using mock data with a known correlation. We generate an ensemble of $M$ arrays of complex gaussian variates $\vecb{x} = (x_0, \ldots, x_{N_t})$, each with a length $N_t$, and convolve each with a normalized gaussian function to introduce a correlation. We then take progressively longer averages of the absolute square of these arrays to mimic the form of the power spectrum estimator \eqref{eqn:pspec_est}.
\begin{equation}
q(l) = \frac{1}{l} \sum\limits_{j=1}^{l} |x_j|^2 \qquad \text{ for } l \leq N_t
\end{equation}

For each averaging length $l$, this gives us an ensemble of $M$ averages $\lbrace q(l) \rbrace$. We can fit a scaled $\chi^2$ distribution to this set using maximum likelihood estimation. The distribution function has the form
\begin{equation}
    f(q; k, \alpha) = \frac{(\alpha q)^{k/2-1} \exp(-(\alpha q)/2)}{2^{k/2} \Gamma(k/2)}
    \label{eqn:chi2}
\end{equation}
where $\alpha$ is a scaling parameter and $k$ is the degree of freedom parameter, and $\Gamma(k)$ is the gamma function. When defined from a sum of squares of unit variance variables, the $\chi^2$ distribution has a mean equal to its degrees of freedom. In the form \eqref{eqn:chi2}, the scaling parameter reflects that $x$ is not unit variance and that $q$ is an average, not a sum.

\begin{figure}
    \centering
    \includegraphics[width=0.45\textwidth]{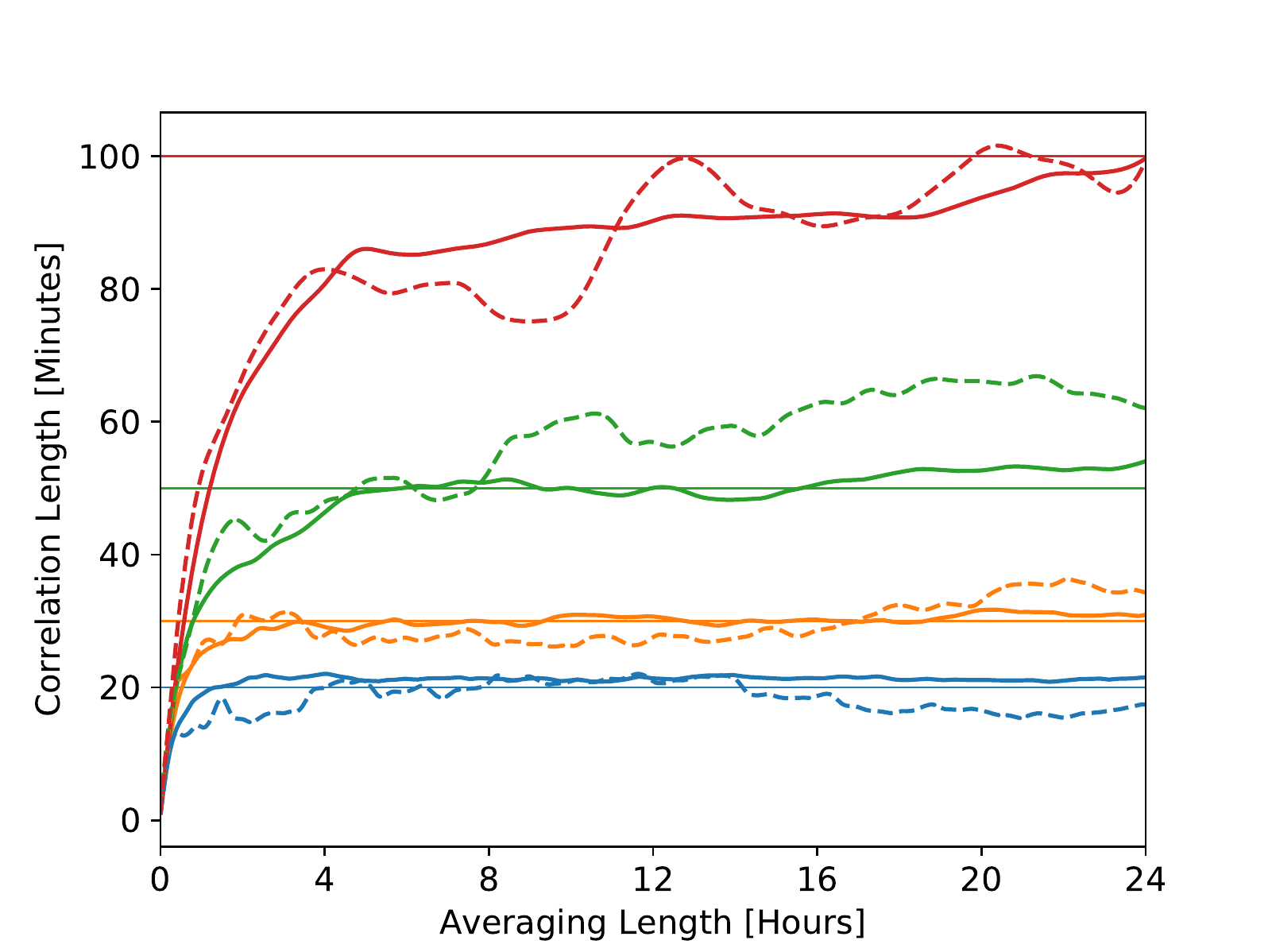}
    \caption{Estimated correlation time vs. averaging length for the toy model of \cref{sec:toy_model}. The correlation time is estimated as in each averaging length by fitting the scaled $\chi^2$ distribution of \cref{eqn:chi2} and using \cref{eqn:tc_est}. The horizontal lines are the true correlation times. The solid lines were made with an ensemble size $M=1000$, the dashed lines are for $M=100$. The correlation times shown are typical of the baseline cases described in \cref{sec:simulation}. The poor convergence for longer correlation times and smaller ensembles indicates that we need to be careful interpreting the results of fitting a $\chi^2$ distribution to power spectrum estimates.}
    \label{fig:corr_convergence}
\end{figure}

We expect that the degrees of freedom in the fits should be related to the correlation time, and we do indeed find that
$$
\frac{l}{k} \rightarrow l_c  \text{ as } l \rightarrow N_t
$$
where $l_c$ is the full-width at half maximum of the convolution kernel. This is shown in \cref{fig:corr_convergence}. The scaling parameter $\alpha$ converges to $\alpha \rightarrow \sigma^2 / (k-1)$, where $\sigma^2$ is the variance of $x$.

To make for a better comparison with our simulations, we define a time step $\Delta t = 11$\,\si{\sec} and use $N_t = 7854$ samples as described in \Cref{tab:sim_params}. The convolution kernel width is set in terms of time as $l_c = t_c/\Delta_t$ for $t_c \in \lbrace$20\si{\min}, 30\si{\min}, 50\si{\min}, 100\si{\min}$\rbrace$. For these cases, we try to recover the input $t_c$ through the estimator
\begin{equation}
\widehat{t}_c = k \: \Delta t
\label{eqn:tc_est}
\end{equation}
where $k$ is the fitted degrees of freedom. \Cref{fig:corr_convergence} shows the result.

For the simple, short-range correlations of the delay-transformed visibilities we can still expect the $\widehat{P}_N$ to be $\chi^2$ distributed, and that we will be able to relate the correlation time. However, the convergence of $\hat{t}_c$ to the true value is slow, and depends on the number of samples going into each fit and how the correlation time compares with the total time available (24 hours). For longer correlation times and fewer samples, the convergence is worse.

Overall, we interpret the results of this toy model to mean that for the baseline types, beam widths, and averaging lengths under consideration, the distribution of $\widehat{P}_N$ will be best described by a $\chi^2$ distribution with relatively few degrees of freedom. The degrees of freedom are connected to the correlation time, but it may not be possible to reliably estimate $t_c$ from fitting the degrees of freedom. For the rest of this work, we use more robust methods to estimate the correlation time and compare with the results of a $\chi^2$ fit for consistency.


\section{Simulation}
\label{sec:simulation}

Simulations were carried by directly evaluating the radio interferometry measurement equation in sky coordinates \citep{thompson_interferometry_2017}, using an analytic and frequency-independent primary beam function. The simulator code is publicly available.\footnote{https://github.com/RadioAstronomySoftwareGroup/healvis} We consider the ideal case of no foregrounds or thermal noise, and that our baselines perfectly measure the Stokes-I visibilities without any polarization leakage. By making these ideal assumptions, we can examine the effects of sample variance in isolation.

We integrate the measurement equation numerically across a spherical sky with uniform pixel areas $\omega$.
\begin{align}
V_b(\nu, t_i) &= \int\limits_\text{FoV} T(\hat{s}, t_i, \nu) A(\hat{s})
e^{-2\pi i \hat{s} \cdot \vecb{b} \nu/c} d\Omega \nonumber\\
&= \sum\limits_{n\in \text{FoV}}  \omega T(\hat{s}_n, t_i, \nu) A(\hat{s}_n)
e^{-2\pi i \hat{s}_n \cdot \vecb{b} \nu/c} \label{eqn:sim_int}
\end{align}

The field of view (FoV) in \cref{eqn:sim_int} is a circular selection of the sky centered on the zenith extended out to a chosen angular distance. For foreground simulations, this should be set to extend to the horizon, (see \citealt{thyagarajan_foregrounds_2015}), but since the choice of FoV increases the number of points to be integrated (especially near the horizon) increased FoV comes at the expense of increased runtime. We choose the FoV diameter to be $110^\circ$, at which distance the widest beam used in this analysis drops to 0.01\% of its maximum.

For each time step, the pointing center is calculated using \texttt{astropy} coordinate frame transformations \citep{astropy:2013, astropy:2018}. The zenith angle and azimuth coordinates of each pixel in the FoV are calculated relative to this pointing center, ensuring that the sky rotates correctly throughout the simulation. Each pixel can therefore be thought of as small, uniform-brightness patch whose positions are in the international celestial reference frame (ICRS). This integration scheme ignores variation in the beam and fringe across the pixel, treating the beam and fringe as having a uniform value across each pixel equal to their values at the center of the pixel. Comparisons between simulations of different resolutions showed minimal effect from these assumptions.

We use a simple Gaussian function, $A(\theta, \phi) = \exp\left(-\theta^2/2\sigma^2\right)$ as our primary beam, where $\sigma$ sets the beam width and $\theta$ is the zenith angle. From here on, ``beam width'' refers to the FWHM of this Gaussian function ($\approx 2.355 \sigma$). This is chosen for computational simplicity and to allow us to explore the effects of varied beam width directly. More realistic beam models, especially those obtained from calibrator sources or EM modeling, are important when modeling foregrounds, because bright sources moving through the beam and its sidelobes can introduce a spectral structure \citep{pober_importance_2016}. Since the signal we analyze here is isotropic, there won't be any drastic changes in integrated flux due to, say, a bright source near the edge of beam dropping into a null at lower frequency. We therefore don't expect these subtle beam effects to be important here.

We note that, though well-defined in sky coordinates, this beam function does not directly correspond with a physical antenna on the ground, such that we are ignoring the potential impacts of beam chromaticity. Within the chosen bandpass, the effects of baseline chromaticity and the signal structure are much more important. Future work will refine these results with more realistic, chromatic beams.

\subsection{Mock signal model}

Instrument simulation tools are generally designed to read sky models in instrument coordinates (angle and frequency). Since frequency maps to distance, the survey volume forms a spherical shell with thickness corresponding with the bandwidth. This shell is divided into voxels with perpendicular area $r_\nu^2 \omega$ and thickness $\delta r_\nu = |r(\nu+\delta\nu) - r(\nu)|$, where $r_\nu$ is the comoving distance corresponding with frequency $\nu$, $\omega$ is the pixel solid angle, and $\delta\nu$ is the frequency channel width. Within the HEALPix pixellization scheme, the \texttt{Nside} parameter sets the pixel solid angle to $\omega = 4\pi/(12 \times $\texttt{Nside}$^2)$ \si{\steradian}, such that all pixels have the same area and pixel centers are evenly spaced \citep{gorski_healpix_2005}.

For our sky model, we select a brightness for each voxel sampled from a Gaussian distribution with frequency-dependent variance $\sigma^2(\nu)$. A space with equal voxel volumes has a flat power spectrum with amplitude $\sigma^2 dV$, since there are no spatial correlations among the values. The comoving voxel volume changes with distance for two reasons. The radial thickness of the voxel changes because the frequency channels are evenly spaced and are nonlinearly related to comoving distance. In addition, the transverse comoving size of the voxel grows quadratically with distance because the solid angle remains the same. To account for this, we scale the variance at each frequency to ensure that the power spectrum amplitude is the same across the band.
\begin{align}
\sigma^2(r) = \frac{\sigma_0^2 dV_0}{dV(r)}
\end{align}
Here, $dV(\nu) = r_\nu^2 \omega \delta r_\nu$ is the comoving voxel volume as a function of frequency. Quantities subscripted $0$ refer to a reference frequency, chosen to be the center of the bandpass.

The sky model used here has a flat power spectrum, $P(k) = \sigma_0^2 dV_0$, which has some advantages for further analysis, and is simple to generate in large quantities. The problem of generating a more realistic EoR model involves establishing spatial correlations that are isotropic and homogeneous within the spherical shell geometry, and is not trivial. It will be of interest to simulated data from EoR models that take into account both the spherical nature of the sky and the expected non-gaussian features of the late-stage EoR signal, especially since these nonlinearities can increase uncertainty on power spectrum measurements \citep{mondal_effect_2015}. We leave full-sky modeling of EoR signals to future work.

\section{Key Questions}
\label{sec:settings}

We ran three sets of simulations to explore the following questions:
\begin{enumerate}
    \item What is the relationship between correlation time and baseline length / beam width?
    
    We explore the validity of \cref{eqn:delay_vis_covar} by running a suite of short simulations with a variety of baseline lengths and beam widths, and calculate the correlation times from them.
    
    \item How does the RMS error of single baseline delay spectra drop off with increasing averaging time for HERA-like and MWA-like beams?
    
    Here we wish to know how well a single baseline can measure the EoR power spectrum through a delay transform method. 
    
    \item What is the lowest RMS error that can be achieved using all the elements of a 37 element hexagon, like MWA Phase II?
    
    Combining data from multiple independent baselines can improve the sensitivity. Within the basic delay spectrum formalism, we are limited, however, to combining baselines of similar lengths, which correspond to the same $k_\perp$ mode. With all antennas in the 37-element array, how much does this binning improve the RMS error?
    
\end{enumerate}

\Cref{tab:sim_params} summarizes the common parameters used by the simulations. Each simulation used the same bandpass, channel width, and array location. The latitude of the array does affect how baselines move through Fourier space over time, which may impact results. We use the HERA array position for all simulations, which is within 5$^\circ$ of latitude from the MWA site, so we expect our results apply to both cases. We use the MWA channel width, which is the finer of the two instruments, and use a 30\,\si{\mega\hertz} bandpass. Most delay analyses focus on a fiducial 10\,\si{\mega\hertz} band, to avoid redshift evolution effects, but for our purposes choosing a wider bandwidth increases the number of $k_\parallel$ probed.

For our sky models, we choose an \texttt{Nside} of 128 and a band-center pixel brightness variance $\sigma_0^2 = (31$\si{\milli\kelvin}$)^2$, taken from the expected EoR signal at $z \sim 13$, the bottom of our bandpass \citep{pritchard_evolution_2008}. Since we're looking at the relative error, the actual signal amplitude is inconsequential. The choice of \texttt{Nside} is a compromise with available computational resources. The simulator had 120GB of memory available, which was sufficient for a shell with an \texttt{Nside} of 256 (resolution of about $13$\si{\arcminute}) and a 30MHz bandpass, or an \texttt{Nside} of 512 (resolution $6.8$\si{\arcminute}) with a 10MHz bandpass. Both higher resolutions come at the expense of significantly longer run times, and since many tasks involved running hundreds of simulations we choose a modest resolution of \texttt{Nside}=128, which corresponds with an angular resolution of about 27.8\si{\arcminute}. This is still small relative to the smallest interferometric fringes in the simulations, and comparisons with higher resolution simulations showed no noticeable impact on results.

\begin{table}
\begin{center}
\begin{tabular}{|c|c|}
\hline 
$\sigma_0$ & 31mK \\
\hline
Bandpass & 100 -- 130 MHz \\ 
\hline 
Channel Width & 78.125 kHz \\ 
\hline 
Duration & 24 Hours \\ 
\hline 
Integration Time & 11 seconds \\ 
\hline 
Latitude & -30.72$^\circ$ \\ 
\hline 
Longitude & 21.428$^\circ$ \\ 
\hline 
MWA beam FWHM & 31.5$^\circ$ \\ 
\hline 
HERA beam FWHM & 12.13$^\circ$ \\ 
\hline
Field of View & 110$^\circ$\\
\hline
\end{tabular}
\end{center}
\caption{Summary of simulation parameters.}\label{tab:sim_params}
\end{table}

\section{Results}
\label{sec:results}

\subsection{Correlation time vs. Baseline and Beam Width}
\label{sec:cov_meas}

The first set of simulations was done to explore the effect of baseline length and beam width on visibility covariance. For 75 beam widths ranging from 15$^\circ$ to 50$^\circ$ we simulated an array of 50 coplanar baselines pointing East, with lengths evenly spaced from 20 to 100 \si{\meter}. These simulations were repeated for 50 separate sky realizations, and run for 2 hours worth of simulated time.

For each simulated dataset, we delay-transform the visibility array and compute the covariance by cross-multiplying and averaging over delay. By treating the delays as an ensemble axis this way, we implicitly assume that all visibilities follow the same distribution at all delays, which we will show later to be a fair assumption due to the flatness of the power spectrum. This cross-multiplication gives us an $N_t \times N_t$ covariance matrix which is symmetric across both diagonals. We then bin the covariance matrix in lag, producing a correlation function that has a nearly Gaussian shape.

\begin{figure}
    \centering
    \includegraphics[width=0.47\textwidth]{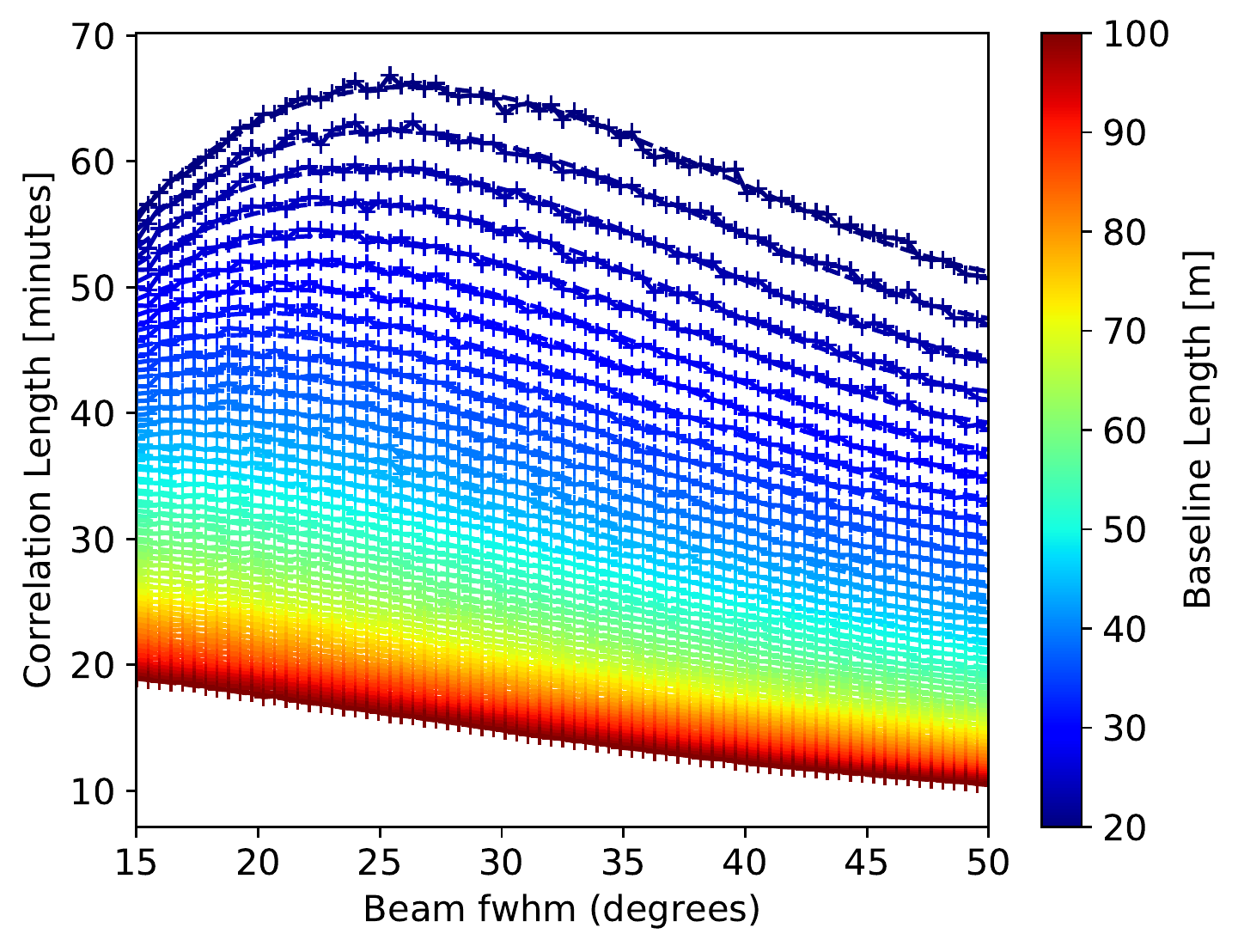}
    \caption{Correlation times vs baseline length and primary beam width for 2 hour simulations, defined as the FWHM of the peak across the diagonal of the covariance matrix.}
    \label{fig:corrlens}
\end{figure}

\Cref{fig:corrlens} shows the results of this set of simulations. As expected from \cref{eqn:delay_vis_covar}, correlation time decreases rapidly with increasing baseline length. For most baseline lengths, correlation time generally decreases with beam width. However, for beam widths less than 25$^\circ$ for baselines shorter than 50 \si{\meter}, correlation time \emph{increases} with beam width. This is unexpected, and counter to our interpretation in \cref{sec:vis_covar}. This is also the regime in which we find the shortest HERA and MWA baselines.

There are several factors at play that may be causing this change in behavior at short baselines and narrow beams. As discussed in \cref{sec:vis_covar}, if we ignore the fringe term in  \cref{eqn:delay_corr_sky} then we would expect the correlation time to increase with beam width. The fringe width is determined by the baseline length, and for short baselines will grow more slowly with time than for a long baseline. It may be that, for short baselines and narrow beams, the decoherence due to narrow beams has a greater effect than the fringe term.

Another possible reason for this turnover may come from our use of a Gaussian primary beams. For a physical antenna, the beam kernel width is approximately equal to the antenna diameter in wavelengths. A Gaussian beam has a Gaussian beam kernel, which does not have a finite width, and therefore would correspond with an antenna of infinite diameter and a diminishing response toward its edge. For short baselines, it may be that the antennas are effectively overlapping under this assumption. However, simulations with more realistic MWA and HERA beam models have hinted at this behavior on short baselines, and so we do believe this is a real effect.

The main takeaway here is that our intuition for correlation, as the overlap of beam kernels in Fourier space, typically leads to the behavior we expect: longer baselines and wider beams give us a shorter correlation time, and so visibilities will decorrelate faster with time and sample variance will drop off faster. However, the shortest baselines in HERA and the MWA do fall into the counter-intuitive regime where increasing beam width increases correlation time. We will recall this behavior on short baselines when interpreting the results of the 24 hour simulations exploring sample variance dropoff.

\subsection{Sample Variance}
\label{sec:rms_err_vs_t}
The second set of simulations was done to measure the root-mean-square (RMS) error of the power spectrum estimator \eqref{eqn:pspec_est} for a range of observation times for individual HERA- and MWA-like baselines. By ``observation time,'' we mean the total contiguous simulated time that is averaged together in the estimator \cref{eqn:pspec_est}. The RMS error $s_k(t)$, with different observation times $t$, as a fraction of power spectrum amplitude, for each $k=k_{b \tau}$ mode, is defined as
\begin{equation}
s_k(t) = \frac{\sqrt{\avg{|\widehat{P}(k; N_t) - P_\text{theory}|^2}}}{P_\text{theory}}
\label{eqn:rms_meas}
\end{equation}
where $P_\text{theory} = \sigma_0^2 dV_0$ is the input power spectrum amplitude.

The MWA/HERA beam widths used here are set based on the true HERA/MWA beam widths at the lowest frequency of the selected bandpass, where they are widest. This is a conservative approximation, since the covariance is smaller for wider beams, so any error from the lack of beam chromaticity is likely to cause us to underestimate sample variance. The HERA beam width is set to $12.13^\circ$, which is the FWHM of an Airy disk pattern for a circular aperture of diameter 14.6\si{\meter} (the HERA dish diameter) at 100\si{\mega\hertz}. Estimating a good effective beam width for the MWA is less straightforward, since each MWA tile is a phased array of 16 dipole antennas and so has a more complicated response than a parabolic dish. We choose a FWHM of $31.4^\circ$, extrapolating from the $23^\circ$ at 137\si{\mega\hertz} estimated in \cite{neben_hydrogen_2016}.

\begin{figure}
    \centering
    \includegraphics[width=0.45\textwidth]{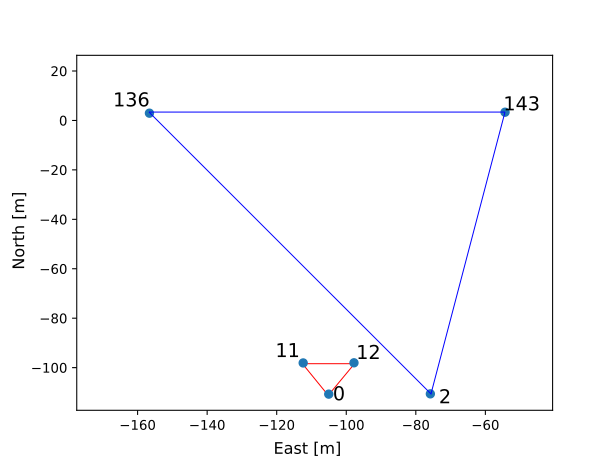}
    \caption{The two baseline configurations, taken from the current layout of HERA.}
    \label{fig:baseline_layouts}
\end{figure}

We simulate 100 sky realizations with each beam width and with two baseline configurations. The first is an equilateral triangle of 14.6\si{\meter} baselines, one oriented East, the others pointing NW and NE. The second configuration is three long baselines: (136,143) points East and is 102.2\si{\meter}, (2,136) points NW and is 139.2\si{\meter}, and (2,143) points NE and is 115.9\si{\meter}. These baselines were selected from the current 65 element layout of HERA, and are plotted in \cref{fig:baseline_layouts}.

For each baseline, we obtain one delay spectrum per time step by doing an inverse FFT along the frequency axis, taking the absolute square, and applying the cosmological scaling factors as in \eqref{eqn:pk_full}. We want to measure the variance of progressively longer averages of these power spectra (i.e., vary $N_t$ in the sum of \eqref{eqn:pspec_est}). We consider a set of observation times $t \in T_\text{avg}$, which range from one time sample (11s) to the full day (24 hours).

To measure the statistics of the estimator, we need an ensemble of averaged power spectra. We build this ensemble by taking separate advantage of all parts of the dataset that independent of each other. For each observation time $t$, we partition the time array into non-overlapping segments of length $t$, such that $N_\text{parts} = \text{floor}(24\text{hrs}/t)$ where the ``floor'' function rounds its argument down to the nearest integer. The single time-step delay spectra are averaged within each segment, giving us $N_\text{parts} \times N_\text{skies}$ independent samples of power spectra averaged at a length $t$, where $N_\text{skies} = 100$ is the number of sky realizations. We then measure the RMS error of each $P(k)$ across this ensemble, giving us the $s_k(t)$ of \eqref{eqn:rms_meas}. This partitioning method naturally gives more samples at shorter averaging time, since there are more short non-overlapping segments within 24 hours than long. Using multiple sky realizations ensures that there are always at least $N_\text{skies}$ samples in the ensemble.

\begin{figure}
    \centering
    \subfloat[\label{shortrms} Three short baselines.]{\includegraphics[width=0.45\textwidth]{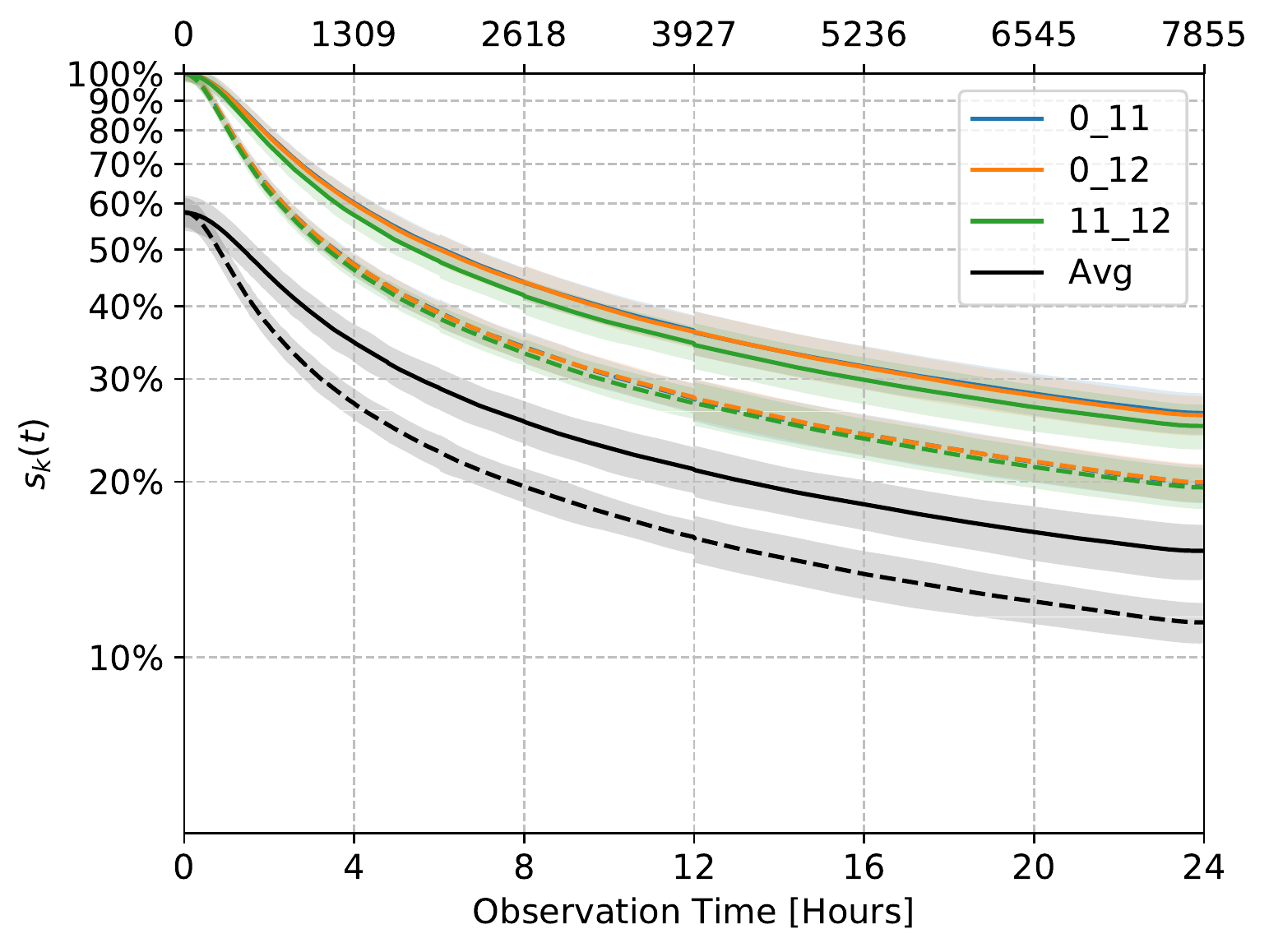}}

    \subfloat[\label{longrms} Three long baselines. ]{\includegraphics[ width=0.45\textwidth]{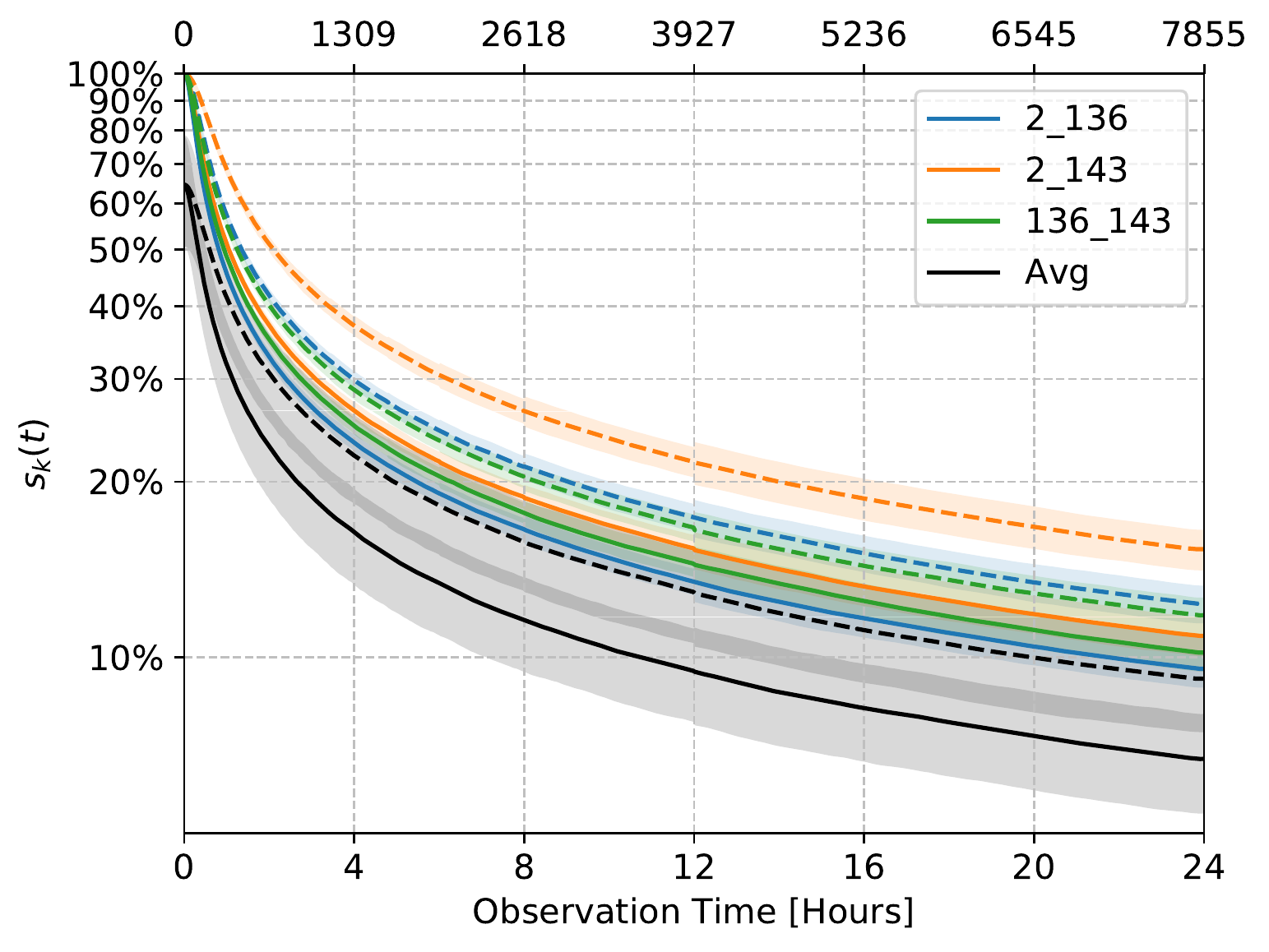}}
    \caption{RMS error vs. averaging time. \protect\subref{shortrms} shows the three 14.6\si{\meter} baselines individually in color, and the result from averaging across the three baselines in black. Dashed lines are for the HERA-like beams and solid lines are for MWA-like beams. \protect\subref{longrms} shows the same results for the three long baselines. Shaded regions are $1\sigma$ errors, estimated across the delay axis. The labels on top indicate the number of integration times.}
    \label{fig:longshort_rmscurv}
\end{figure}

\Cref{fig:longshort_rmscurv} shows the RMS of power spectrum estimates at different time observation times. The lines show the mean of $s_k(t)$ across all $k$, while the error shading show the standard error calculated across $k$.\footnote{Note that this is effectively the variance of the variance of the power spectrum, which is itself a variance.} The discrete jumps in the shaded regions are due to the sharp changes in $N_\text{part}$ at certain observation times.

For the short baselines, the RMS drops off more quickly for the HERA beam (dashed) than for the MWA beam (solid). This is consistent with the conclusions of \cref{fig:corrlens}, since the 14.6\si{\meter} baselines are in the regime where increasing beam width increases correlation time, which means a slower drop-off. This behavior is reversed in the long baseline case. The black lines and grey shading are the corresponding results after binning power spectra in $k_\perp$. Averaging together the three baselines in each case drops the error by a factor of $\sim \sqrt{3}$, as expected for independent baselines.

The baseline length appears to have the most significant effect on RMS error, yet even with a 139\si{\meter} baseline the RMS error doesn't get below $10\%$. The situation is much worse for the short baselines, which individually don't reach below 19\% for HERA beams or below 25\% for MWA beams. Single-baseline estimators, particularly for short baselines, cannot be used reliably for precision cosmology.

\subsection{Full array simulations}
\label{sec:full_array_simulations}

Results for individual baselines of specific lengths and orientations motivated us to determine the sample variance limits of a 37 element hexagonal array, i.e., the layout of the MWA Phase II East hexagon. Antenna elements are spaced by 14\si{\meter} in this configuration. Baselines are grouped by redundancy and one baseline is selected from each redundant group. We ran a series of simulations of 100 sky realizations, calculating the visibilities only for these selected baselines. Since there is no thermal noise in this simulation, visibilities from redundant baselines are identical. For the 37 element hexagonal array, there are 66 unique baseline types.

\begin{figure*}
\centering
\includegraphics[width=1.0\textwidth]{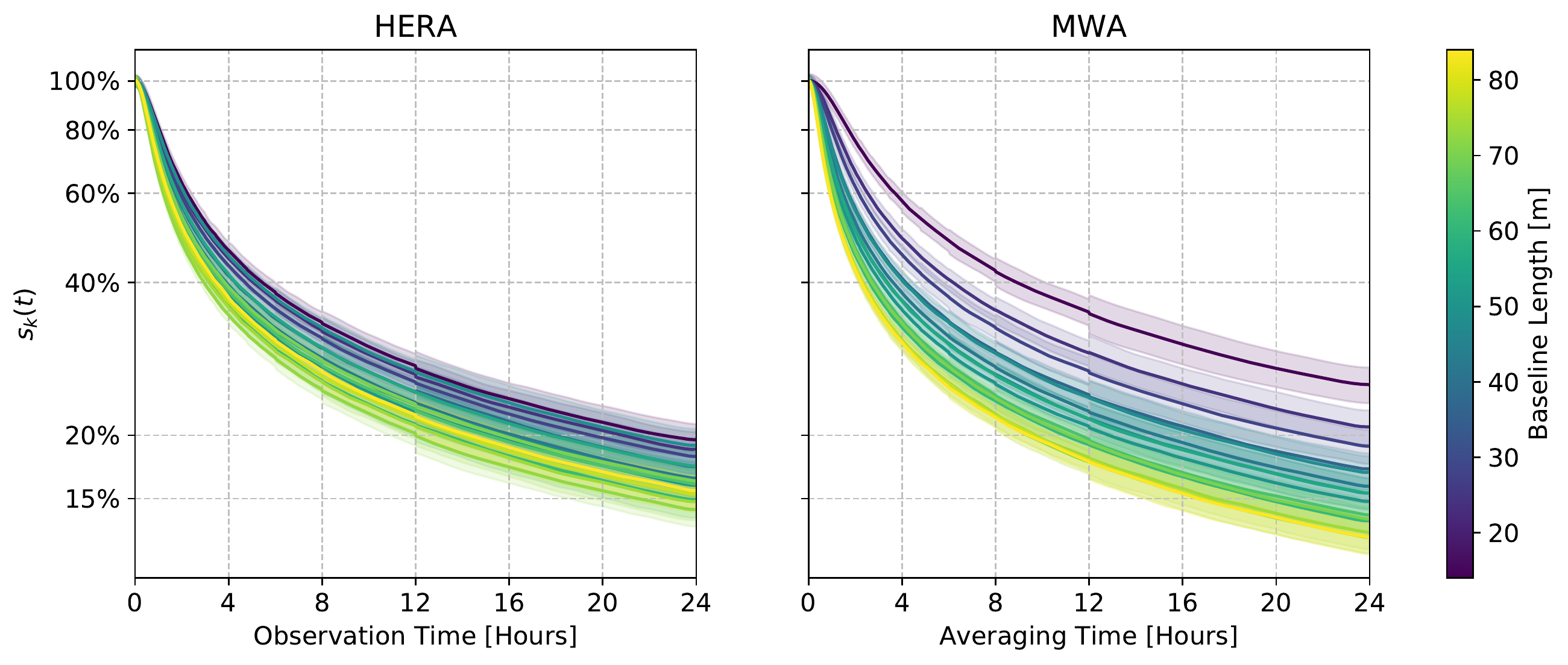}
\caption{RMS Error vs averaging time for the unique baseline lengths in the 37-element hexagon simulations.}
\label{fig:rms_reds_perbl}
\end{figure*}

Each baseline samples the same range of $k_\parallel$ for a given $k_\perp$. \Cref{fig:rms_reds_perbl} shows the RMS errors of single-baseline power spectrum estimates for the different baseline lengths in the array vs. averaging time. The discontinuities at 12 hours, 8 hours, and elsewhere are due to the sudden changes in the number of partitions available.

The spread of curves for the MWA beam, compared with the HERA beam curves, is due to the change in behavior observed in \cref{fig:corrlens}: for short baselines the narrower HERA beam has a shorter correlation time than the MWA, and vice versa for long baselines. The general trend of longer baselines having faster variance dropoff is clear for both instruments.

We can see from \cref{fig:rms_reds_perbl} that, for individual baselines in the 37-element hexagonal layout, $s_k$ stays above about 13\% for MWA beams and above 15\% for HERA beams. This is in the optimistic case that all 24 hours of LST can be used. With only 8 hours of data, which is how much time was used by \cite{ali_paper-64_2015} for PAPER-64, the error for a single baseline of HERA is around 35\%, and is upwards of 45\% for the MWA.

\begin{figure}
    \centering
    \includegraphics[width=0.45\textwidth]{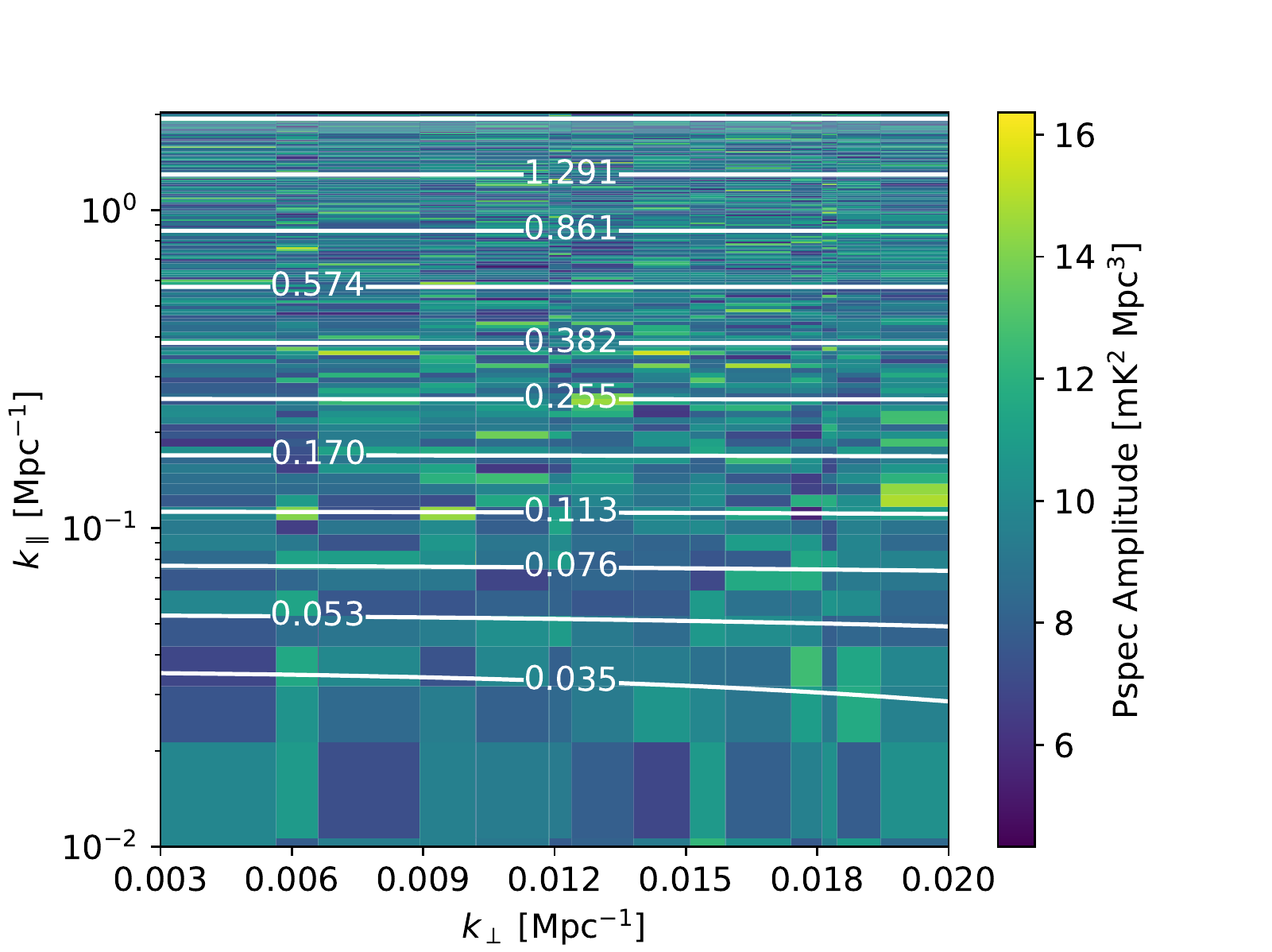}
    \caption{The cylindrical power spectrum from a HERA beam simulation with a 37 element hexagonal array, with contours indicating the boundaries of $|k|$ bins used in \cref{sec:full_array_simulations}. The full range of $k_\parallel$ goes up to about 2 Mpc$^{-1}$, but the contours get crowded at high $k_\parallel$ on this log-scale and so we only plot up to 0.6 Mpc$^{-1}$. For most $k_\parallel$, the spherical-binning effectively combines all $k_\perp$ in the array.}
    \label{fig:cylspec_contour}
\end{figure}

By combining estimates from multiple baseline lengths, we can reduce this error dramatically. We can extend \cref{eqn:pspec_est} by binning separate $k_{b,\tau}$ modes. Each $k_{b,\tau}$ is a single delay mode for a single baseline, and coresponds with a point in the cylindrical power spectrum space $(k_\perp, k_\parallel)$. We can define bin in the spherical-averaged $k = \sqrt{k_\parallel^2 + k_\perp^2}$, where $k_\parallel = \tau/Y$ and $k_\perp = b/(\lambda X)$, and average together $\widehat{P}(k_{b,\tau}, N_t; b)$ in each bin.

Choosing a set of $k$ bins is a subtle problem that depends on the expected shape of the power spectrum. For a highly-structured power spectrum, one needs to choose fine $k$ bins to resolve the structure. For a smoother spectrum, one can combine measurements into wider $k$ bins with little loss of information. Choosing wider bins means more measurements are combined, which can reduce sample variance, so there is a trade-off between the precision with which we can constrain the power spectrum shape and the per-bin sample variance. For a flat power spectrum model, like the ones we're using, one could in principle average all $k$ modes with impunity, since there is no shape to constrain.

We choose to bin in the logarithmically spaced $k$ bins of the default configuration in \texttt{21cmFAST}, which provide a representative range of scales and $k$ space resolution for describing the information in power spectra typical of the EoR \citep{mesinger_21cmfast_2011}. The bin edges are shown as contours in \cref{fig:cylspec_contour}, over the power spectrum $P(k_\perp, k_\parallel)$ averaged 24 hours for a single sky realization.


\begin{figure}
    \centering
    \includegraphics[width=0.45\textwidth]{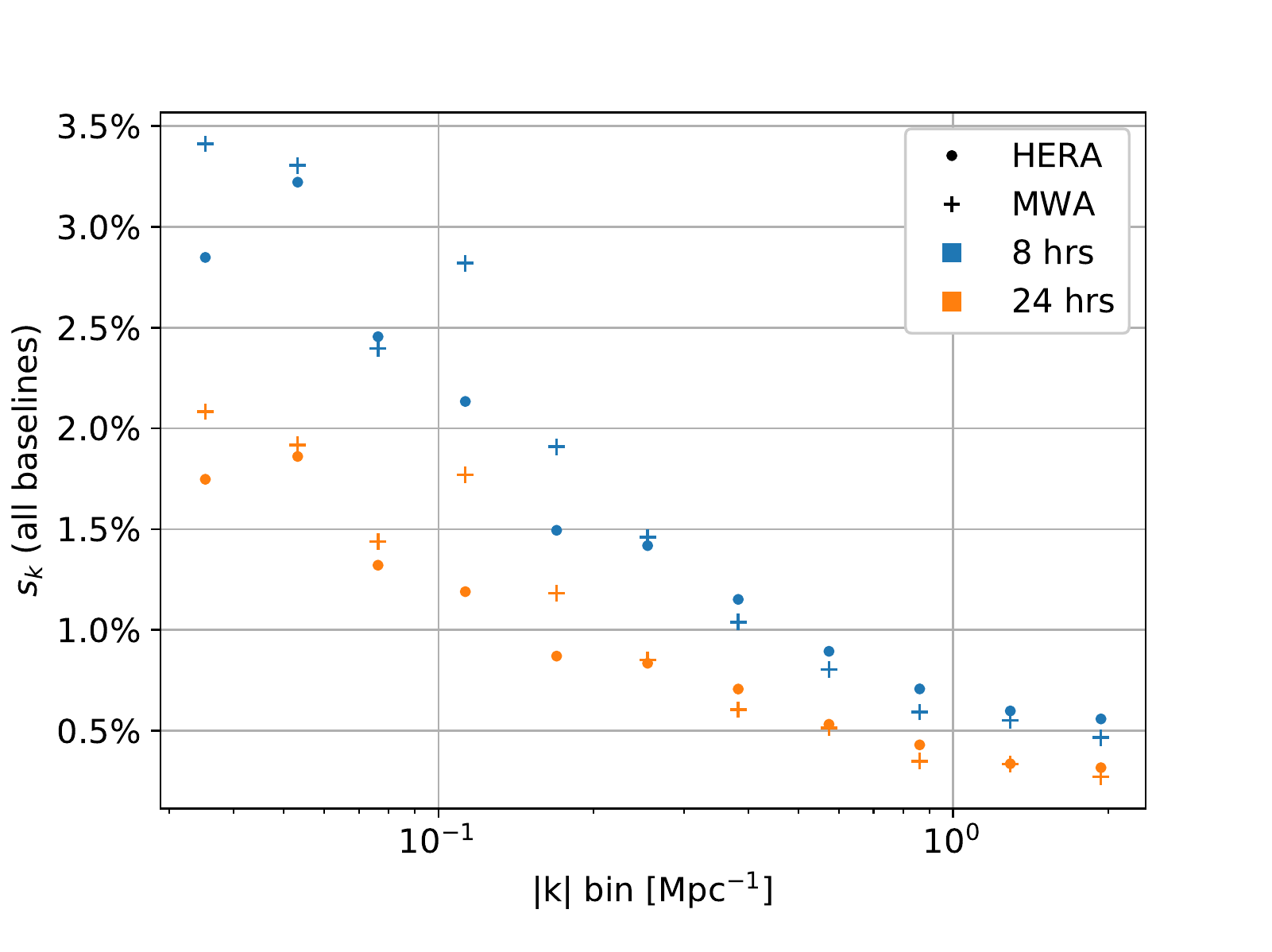}
    \caption{The RMS error of the spherically-binned power spectrum in each logarthmically-spaced $|k|$ bin, including baselines of all lengths in the 37 element hexagonal array. Blue points are the result of averaging 8 hours of LST, and orange are for 24 hours of averaging. Crosses are for the MWA-beam simulations, and circles are for the HERA beam simulation. More modes are included in bins with higher $|k|$ since the bin width increases logarithmically, which results in a reduced sample variance compared with smaller bins at lower $|k|$. For a $k$ binning scheme typical of an EoR experiment, the 37 element hexagonal array brings the sample variance error below 3.5\% for all bins, and can bring it under a percent for some.}
    \label{fig:sph_avg_37limits}
\end{figure}


\Cref{fig:sph_avg_37limits} shows the RMS error in each $k$ bin, at different observation times and for the different instruments, using all baselines in the 37 element array for both HERA and MWA-like beams and both 8 and 24 long observations. The variance drops off with increasing $k$ due to the logarithmic bins, which become wider at high $k$ and therefore include more samples. We note that for all bins considered, the error is under 3.5\%, and can be brought under a percent for the highest bins. Therefore, we expect that even with a relatively small number of antennas, a maximally redundant array can place high precision constraints on the power spectrum, as long as all baseline types are used in the measurement.

\section{Comparing Correlation Time Measures}
\label{sec:res_corrlen}

\begin{figure}
    \centering
    \includegraphics[width=0.45\textwidth]{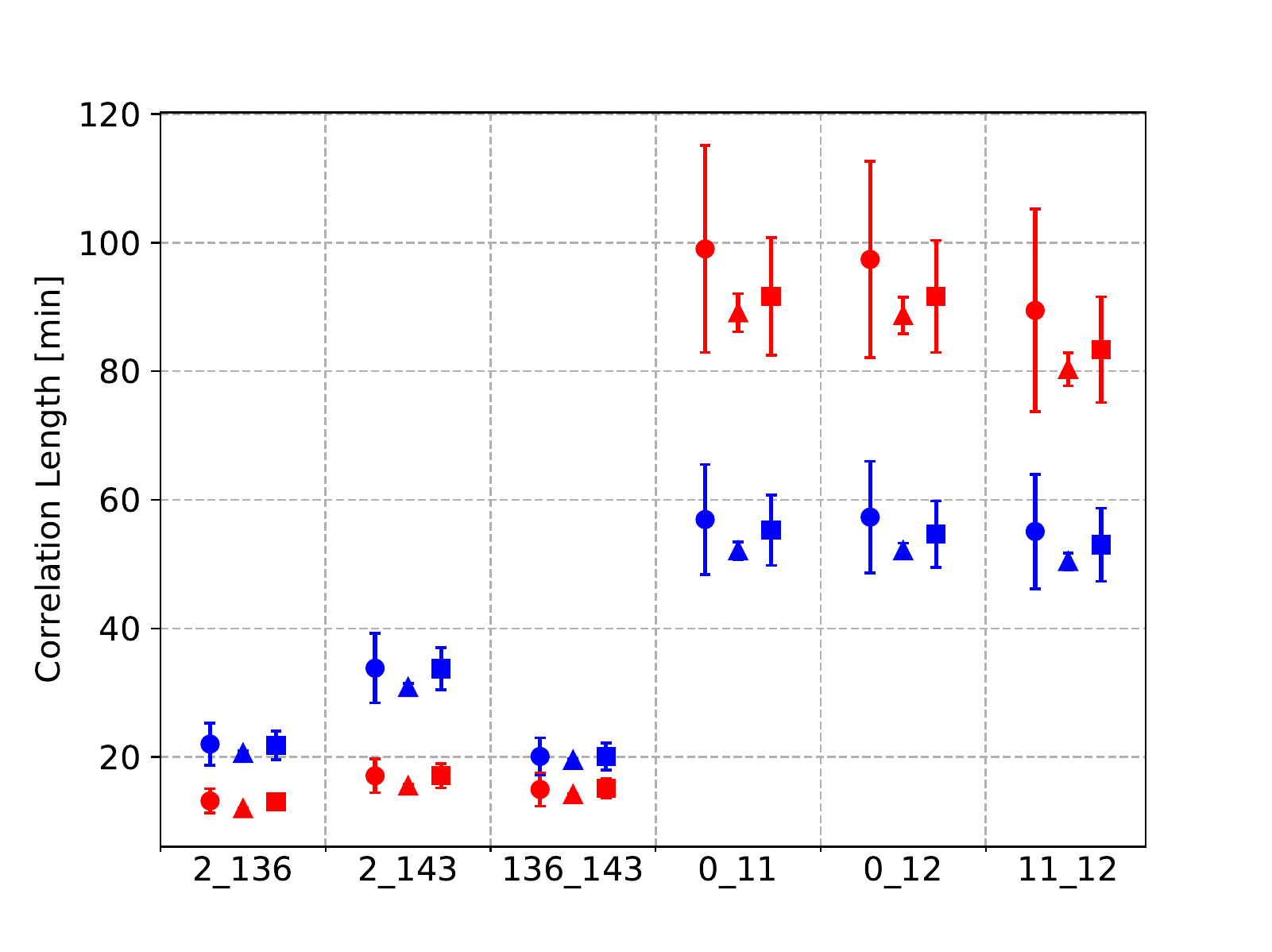}
    \caption{Correlation times measured with three different methods, for the six baselines in \cref{fig:baseline_layouts}. Red and blue are the results for MWA and HERA beams, respectively. Circles, triangles, and squares are results from the 24 hour RMS measurement, the binned covariance matrices, and from fitting $\chi^2$ distributions to data at 12 hours of averaging, respectively (see \cref{sec:res_corrlen}). All three estimates give consistent results.}
    \label{fig:threemethod_corlen_compare}
\end{figure}

The correlation time $t_c$ specifies the approximate separation in time of independent samples of the power spectrum, and can in principle be calculated from the baseline length and primary beam model. This allows one to estimate the sample variance error for a given baseline and beam model directly from a formula like \cref{eqn:variance_function}, without running expensive simulations. In this section, we show that the correlation time as defined in \cref{sec:cov_meas} gives the RMS error for longer averaging times, and is consistent with the number of degrees of freedom in a $\chi^2$ distribution, using the six baselines simulated in \cref{sec:rms_err_vs_t}. We also show that using the correlation time calculated here fits the RMS error vs. time plots, using \cref{eqn:variance_function}. The results are summarized in \cref{fig:threemethod_corlen_compare}.

\subsection{Binned covariance}

The first $t_c$ measurement method is to calculate the covariance matrix, bin it in lag, and find the width of the central peak. Since we have $N_\text{skies}$ simulations in each configuration, we can calculate the covariance matrix separately for each delay mode. This is in contrast to the method in \cref{sec:cov_meas}, where we used the delay axis as an ensemble when computing covariance matrices. We switch to this method now because it is more consistent with the per-k RMS error measurement of \cref{sec:rms_err_vs_t}. The results of this method are plotted as triangles in \cref{fig:threemethod_corlen_compare}. The points and error bars are the means and standard deviation, across $k$, of the measured correlation times.

We note that there is no technical reason why the FWHM of the covariance should set the correlation time. The correlation beyond the half-maximum distance is not negligible. For this reason, the FWHM is likely underestimating $t_c$. This explains why the triangles in \cref{fig:threemethod_corlen_compare} are a bit lower, though still within the uncertainty, of the results from the other methods.

\subsection{RMS in the central limit}

For a large number of independent samples, the estimator $\widehat{P}_t$ is approximately Gaussian-distributed under the central limit theorem. We therefore expect RMS of the power spectrum estimator should go roughly as $s_k(t) = 1/\sqrt{M(t)}$, where $M$ is the number of independent time samples within the time $t$, for $t >> t_c$. From $M$ we can estimate $t_c$ as
\begin{subequations}
\begin{align}
    t_c \approx t / M = t \left[ s_k(t) \right]^2 \label{eqn:tc_from_rms} \\
    \Delta t_c \approx 2 t s_k(t)  \Delta s_k(t) \label{eqn:tc_from_rms_err}
\end{align}
\end{subequations}
\Cref{eqn:tc_from_rms_err} is the error on the estimate, propagating from the measured error $\Delta s_k(t)$. The squares in \Cref{fig:threemethod_corlen_compare} show this estimate from the 24 hour averaging time, using the results in \cref{fig:longshort_rmscurv} with \cref{eqn:tc_from_rms,eqn:tc_from_rms_err}.

\subsection{$\chi^2$ distribution fit}

The third method, discussed in \cref{sec:toy_model}, is to fit a scaled $\chi^2$ distribution to the measured power spectra and treat the fitted degrees of freedom parameter $\kappa$ as the number of independent samples in the measurement, such that the correlation time is $t_c = t_\text{avg} / \kappa$\footnote{To address a possible confusion in terminology: In most cases, the phrase ``$\chi^2$ fitting'' refers to fitting model parameters to data using a chi-squared goodness-of-fit test. Here, we mean we are fitting the parameters of a $\chi^2$ distribution that describes the statistics of our data.} We take the ensemble of $\widehat{P}_N$ measurements for a chosen observation time, and estimate the parameters of the scaled $\chi^2$ distribution separately for each $k$-mode using a maximum likelihood estimator. The square points on \cref{fig:threemethod_corlen_compare} show the mean of these per-k estimates, with error bars given by the standard error across $k$.

\begin{figure}
    \centering
    \includegraphics[width=0.45\textwidth]{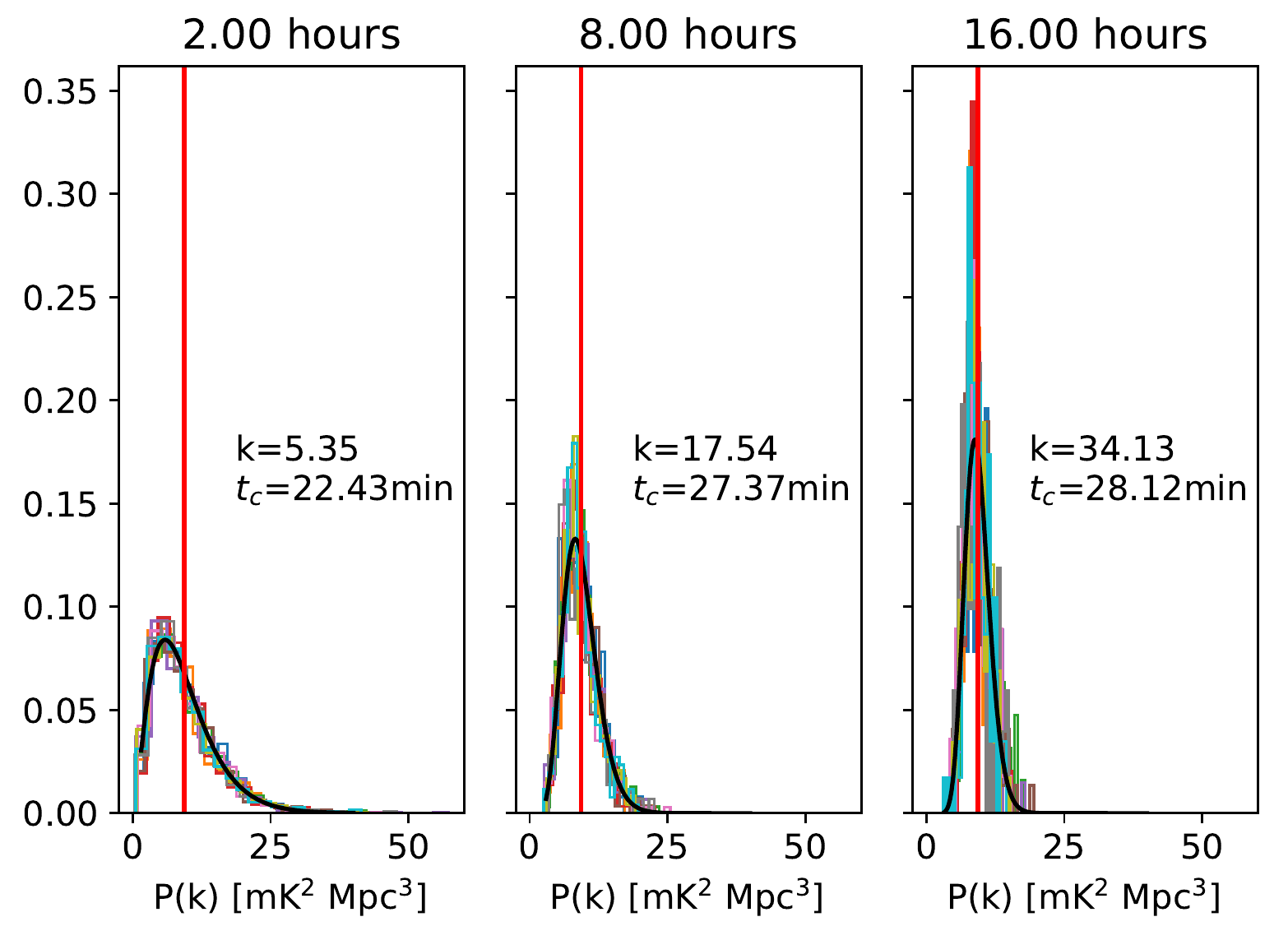}
    \caption{Histograms of the power spectra at each $k$ for the three short HERA baselines configuration, at different observation times. The vertical red line is the theoretical power spectrum amplitude. Each k-mode has approximately the same $\chi^2$ distribution, with effective degrees of freedom increasing with increasing observation time. Note that for longer averaging lengths the ensemble is smaller, which causes the $P_k$ measurements to be noisier.}
    \label{fig:perk_dists}
\end{figure}

Two factors influence how well this method may be used. For shorter averaging times, the ensemble of $\widehat{P}_N$ measurements is much larger, making fewer samples to fit with the $\chi^2$ distribution. However, as \cref{fig:corr_convergence} shows, for longer correlation times (e.g. the red curve), there is a risk of underestimating in this case. With much longer averaging times, there is the opposite problem. Since the other methods suggest the correlation times to be between 20 and 100 minutes, we look at estimates from 12 hours of averaging. \Cref{fig:perk_dists} shows histograms of the averaged power spectra for three observation times, along with the fitted $\chi^2$ for that time. With longer averaging times, the degrees of freedom increases and the distributions become less skewed, as expected. We note that there is still a noticeable skew around 8 hours.

\begin{figure}
    \centering
    \subfloat[\label{shortrmstheory}]{\includegraphics[width=0.45\textwidth]{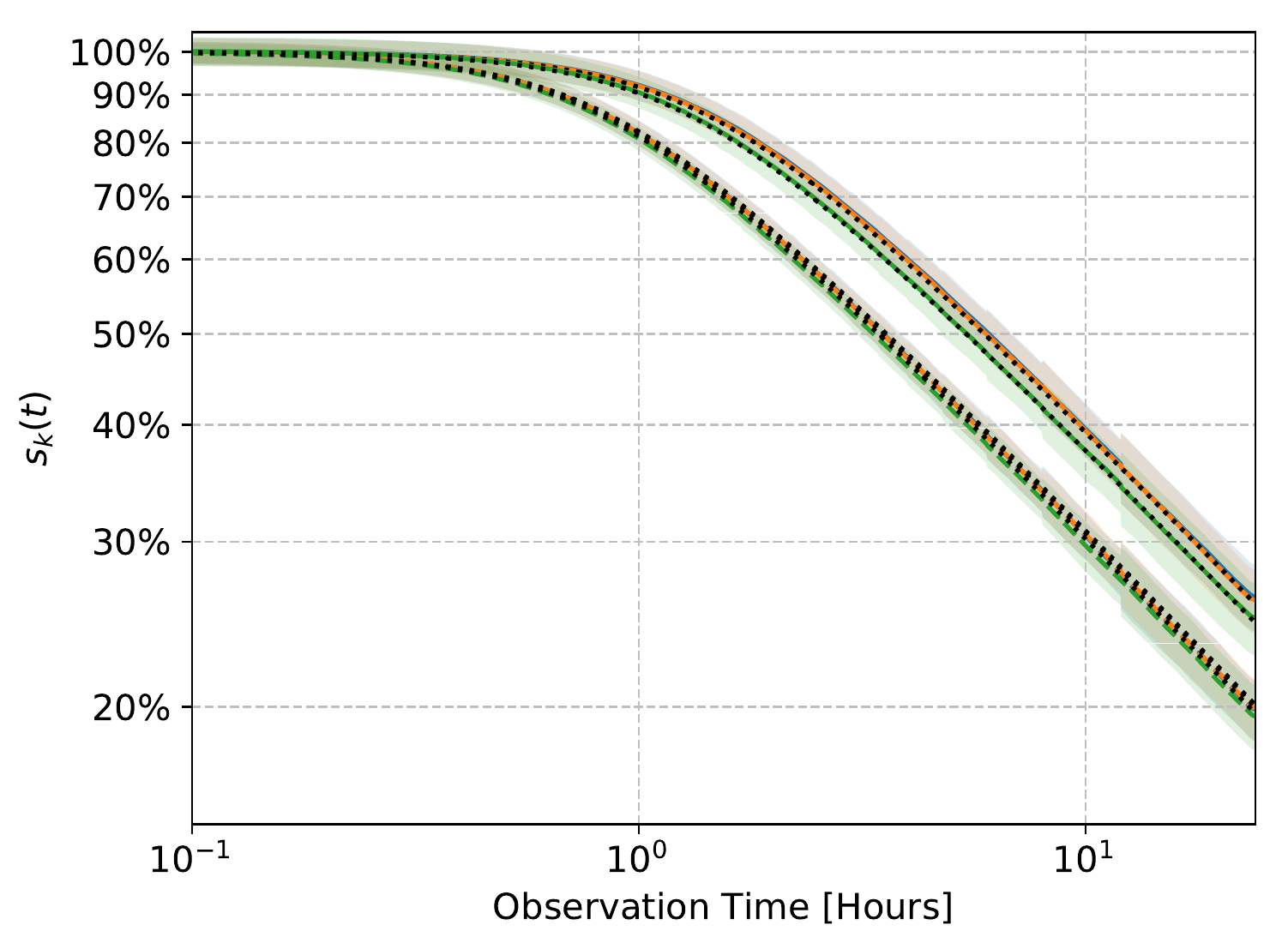}}

    \subfloat[\label{longrmstheory}]{\includegraphics[ width=0.45\textwidth]{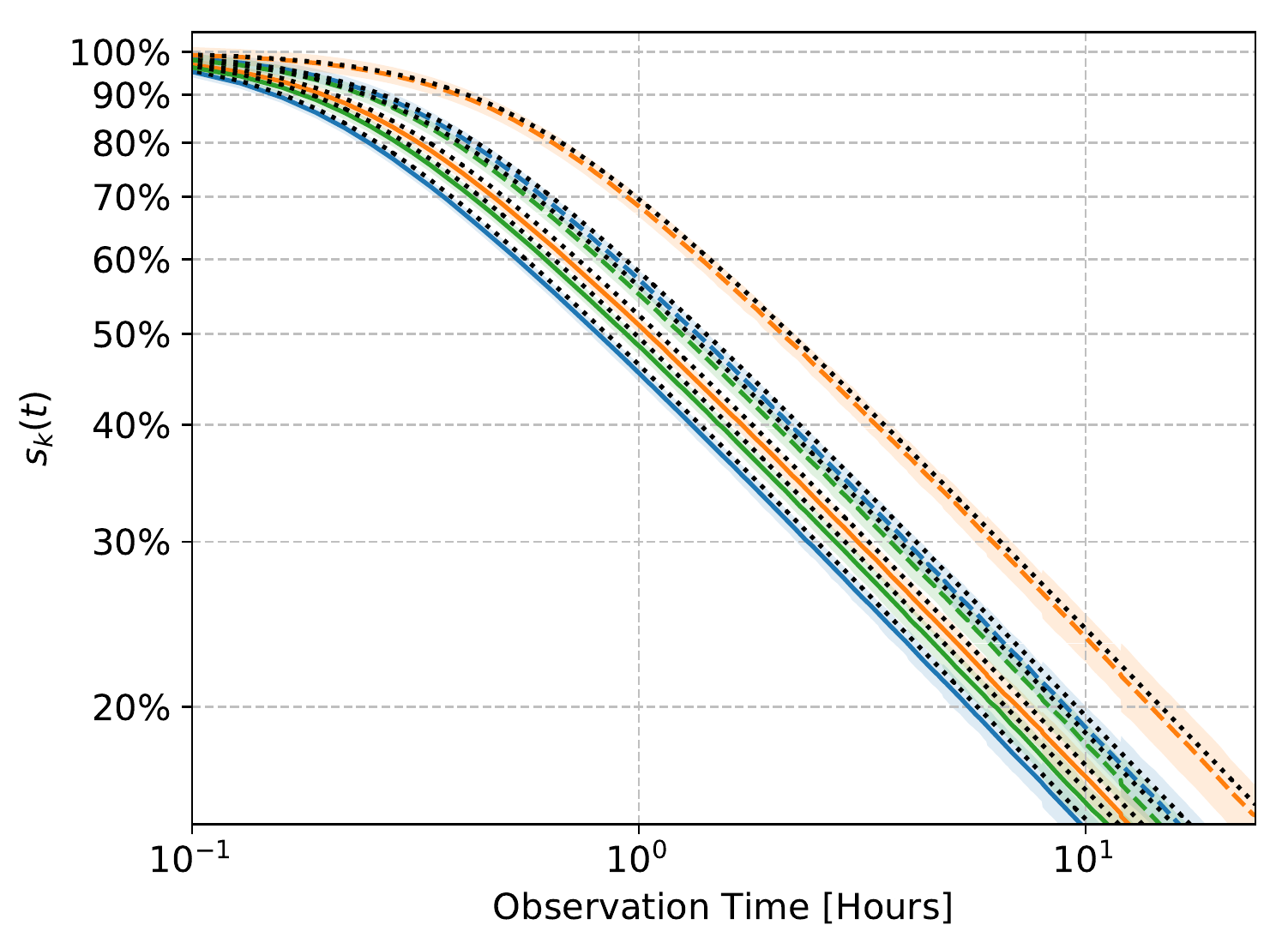}}
    \caption{ The same rms curves as \cref{fig:longshort_rmscurv}, shown with the theoretical RMS curves of \cref{eqn:variance_function} using the correlation times measured from the $\chi^2$ distribution fitting (dashed black lines). The axes are shown in log-log scale to emphasize the flattening at short averaging time.}
    \label{fig:longshort_rmscurv_theory}
\end{figure}

\subsection{Theoretical RMS error}

\Cref{fig:longshort_rmscurv_theory} shows the RMS vs. averaging time curves of \cref{fig:longshort_rmscurv}, but with the theoretical sample variances for those baselines indicated with dashed black lines. This theoretical curve is obtained from \cref{eqn:variance_function} by dividing out by $\gamma_0 = P_\text{theory}$ and taking the square root. The $t_c = 2.355 w_c$ used here are the correlation times measured by $\chi^2$ distribution fitting (i.e. the squares in \cref{fig:threemethod_corlen_compare}).

The good agreement shown in \cref{fig:longshort_rmscurv_theory} indicates that the form of \cref{eqn:variance_function} can be used to estimate the single-baseline sample variance at different averaging times, knowing only $t_c$, in the case that the correlation function is approximately Gaussian in time. The correlation time may be calculated directly from \cref{eqn:delay_corr_sky} or \cref{eqn:delay_vis_covar}. It is thus possible to estimate the sample variance of an experimental setup without carrying out simulations. This may prove invaluable for future instrument design.

\section{Discussion and Conclusions}
\label{sec:discussion}

We have used wide-field instrument simulations with Gaussian beams to estimate the statistical independence of delay-transformed visibilities over time, and calculate the uncertainty per-baseline for the simple delay-spectrum estimator of \cref{eqn:pspec_est}. We have further looked further at the case of a 37 element hexagonal array, like the MWA Phase II, and measured the uncertainty of spherically-binned power spectrum estimates. This was done under the assumption that thermal noise and foreground power have been fully mitigated, and so represents a fundamental limit on the uncertainty that this estimator can reach.

\begin{table*}
\centering
\begin{tabular}{c|c|c|c|c|}
\cline{2-5}
 & \multicolumn{ 2}{c|}{\textbf{MWA}} & \multicolumn{ 2}{c|}{\textbf{ HERA}} \\
 \cline{2-5}
 & \multicolumn{1}{c|}{\textbf{8 hrs}} & \multicolumn{1}{c|}{\textbf{ 24 hrs}} &
 \multicolumn{1}{c|}{\textbf{ 8 hrs}} & \multicolumn{1}{c|}{\textbf{ 24 hrs}} \\ \hline
\multicolumn{1}{|r|}{\textbf{14.6m}} &  43.86\% $\pm$ 2.94\% &  26.22\% $\pm$ 2.13\% &  34.01\% $\pm$ 2.15\% &  19.89\% $\pm$ 1.50\% \\ \hline
\multicolumn{1}{|r|}{\textbf{140m}} &  16.55\% $\pm$ 0.92\% &  9.57\% $\pm$ 0.69\% &  21.23\% $\pm$ 1.12\% &  12.37\% $\pm$ 0.92\% \\ \hline
\multicolumn{1}{|r|}{\textbf{37hex}} & 3.41\% & 2.04\% & 3.22\% & 1.86\% \\ \hline
\end{tabular}
\caption{Results at 8 hours and 24 hours for MWA and HERA beams, for single baselines, and for the spherically-binned power spectra of the 37 element hex. The binning used is discussed in \cref{sec:full_array_simulations}. The maximum uncertainty across all $k$ bins is shown for the 37hex binned case, so no uncertainty is quoted.}
\label{tab:final_results}
\end{table*}

\Cref{tab:final_results} summarizes sample variance estimates in several of the simulated scenarios. The shortest baselines in HERA and the MWA cannot bring the sample variance below 20\%, even in the ideal case that all 24 hours of LST are available. Spherically averaging power spectrum estimates across all baselines in the 37 element hexagonal array brings the RMS error under 3.5\% for all cases.

As mentioned in \cref{sec:full_array_simulations}, choosing spherical-k bins for the power spectrum estimator does come with its own difficulties. In the case of a flat power spectrum, the only parameter we have to constrain is the amplitude, which does not depend on $k$. For a non-flat spectrum, however, binning in $k$ can potentially smooth over small scale features, which will propagate through to constraints on the model parameters. We chose the logarithmic bins of \texttt{21cmFAST} as a typical binning used in 21cm analysis, but the actual choice of $k$ bins will affect the available improvement in sample variance.

We note also that these results largely relied on the Gaussianity of the tested signal. The EoR signal, especially at late times, is expected to be highly non-Gaussian due to the evolution of structure and appearance of ionized regions in the field. \cite{mondal_effect_2015}, and more recently \cite{shaw_impact_2019}, have shown that higher-order statistics such as the trispectrum enhance the errors of power spectrum estimates by  50\% -- 100\% at redshifts up to $z \lesssim 10$. Future simulations with a more realistic EoR model will be able to examine the effect of these non-Gaussianities on realistic power spectrum estimates.

Nonetheless, our results bode well for HERA and the MWA's usefulness for precision cosmology. As suggested by \cite{greig_21cmmc:_2015}, getting the sample variance under 25\% can be enough to constrain EoR parameters within a factor of a few, though the tolerable level of power spectrum uncertainty remains to be determined. As HERA construction continues, improvements in foreground modeling and removal will clean up more of Fourier space, allowing longer baselines to be used. Further, the split-hexagon final layout of HERA, discussed in \cite{dillon_redundant_2016}, will add more non-redundant short baselines to the analysis, providing more independent samples of low $k_\perp$ modes. The MWA phase II also consists of another hexagonal array and a number of randomly-placed tiles, which greatly increases the numbers of available baselines. Most importantly, none of these sample variance uncertainties prevent the possibility of an EoR power spectrum detection, which is the most immediate goal of these instruments.

\vspace{1em}
To summarize our main conclusions:
\begin{itemize}
    \item Sample variance is an important factor on single baselines, and severely limits the precision of cosmological models that can be drawn from power spectrum measurements.
    \item Including measurements from longer baselines can help to mitigate the impact of sample variance, which drops dramatically with increasing baseline length. Combining measurements from multiple baseline lengths by binning across $k_\perp$ will further reduce the sample variance down to tolerable levels, while still avoiding the use of very long baselines.
    \item For a gaussian signal, the uncertainty on a delay spectrum estimator follows a scaled chi-squared distribution. The effective degrees of freedom of this distribution can cause a non-negligble skew, which may further bias parameter constraints by underestimating the error toward larger $P_k$ values. Analyses attempting to make cosmological parameter constraints should account for this potential skew.
    \item The sample variance of a Gaussian or near-Gaussian signal may be estimated from the time-correlation of visibilities. This correlation does not scale simply with beam width and baseline length, but can be evaluated analytically or numerically from knowledge of the primary beam shape and baseline vector.
    \item The time-correlation of delay-transformed visibilities depends on the primary beam width and the baseline length, but can behave in unexpected ways. The correlation time of short baselines will increase with increasing beam width for narrow beams, and decrease for wider beams.
\end{itemize}

\section{Acknowledgements}

This work was supported by National Science Foundation Grant Nos. 1636646 and 1613040, and NASA Grant 80NSSC18K0389. Thanks are owed to  Miguel Morales, Cathryn Trott, Nithyanandan Thyagarajan, Michael Wilensky, Daniel Jacobs, Bryna Hazelton, Adrian Liu, and Bradley Greig for their useful advice and discussions. We thank the developers of \texttt{scipy}, \texttt{matplotlib}, \texttt{astropy}, \texttt{healpy}, and \texttt{pyuvdata}.

\bibliography{sample_var}
\bibliographystyle{mnras}

\label{lastpage}
\end{document}